\documentclass[aps,prb,twocolumn,showpacs,superscriptaddress,longbibliography,notitlepage]{revtex4-2}
\usepackage[colorlinks=true,citecolor=blue,linkcolor=blue,urlcolor=blue,breaklinks=true]{hyperref}
\usepackage{amssymb}
\usepackage{graphicx}
\usepackage{amsmath}
\usepackage{mathrsfs}
\usepackage{mathtools}
\usepackage[export]{adjustbox}
\usepackage{epsfig}
\usepackage{times}
\usepackage{color}
\usepackage{setspace}
\usepackage{calc}
\usepackage{natbib}
\usepackage{bm}
\DeclareMathAlphabet{\mathpzc}{OT1}{pzc}{m}{it}
\usepackage{array}
\usepackage{multirow}
\usepackage{comment}
\usepackage{float}
\usepackage{subfigure}
\usepackage{natbib}

\begin{document}
	
	\newcommand {\beq} {\begin{equation}}
	\newcommand {\eeq} {\end{equation}}
	\newcommand {\bqa} {\begin{eqnarray}}
	\newcommand {\eqa} {\end{eqnarray}}
	\newcommand{\dhat}{\ensuremath{\hat{D}}}
	\newcommand{\ehat}{\ensuremath{\hat{E}}}
	\newcommand{\lhat}{\ensuremath{\hat{\Lambda}}}
	\newcommand{\zbar}{\ensuremath{\bar{\zeta}}}
	\newcommand{\ebar}{\ensuremath{\bar{\eta}}}
	\newcommand {\ba} {\ensuremath{b^\dagger}}
	\newcommand {\Ma} {\ensuremath{M^\dagger}}
	\newcommand {\psia} {\ensuremath{\psi^\dagger}}
	\newcommand {\psita} {\ensuremath{\tilde{\psi}^\dagger}}
	\newcommand{\lp} {\ensuremath{{\lambda '}}}
	\newcommand{\A} {\ensuremath{{\bf A}}}
	\newcommand{\Q} {\ensuremath{{\bf Q}}}
	\newcommand{\kk} {\ensuremath{{\bf k}}}
	\newcommand{\qq} {\ensuremath{{\bf q}}}
	\newcommand{\kp} {\ensuremath{{\bf k'}}}
	\newcommand{\rr} {\ensuremath{{\bf r}}}
	\newcommand{\JJ} {\ensuremath{{\bf J}}}
	\newcommand{\xib} {\ensuremath{{\boldsymbol{\xi}}}}
	\newcommand{\etab} {\ensuremath{{\boldsymbol{\eta}}}}
	\newcommand{\rp} {\ensuremath{{\bf r'}}}
	\newcommand {\ep} {\ensuremath{\epsilon}}
	\newcommand{\nbr} {\ensuremath{\langle ij \rangle}}
	\newcommand {\no} {\nonumber}
	\newcommand{\up} {\ensuremath{\uparrow}}
	\newcommand{\dn} {\ensuremath{\downarrow}}
	\newcommand{\rcol} {\textcolor{red}}
	\newcommand{\bcol} {\textcolor{blue}}
	\newcommand{\bu} {\bold{u}}

	\newcommand{\tr}[1]{\mathrm{Tr}\left[#1\right]}
	\newcommand{\ve}[1]{\boldsymbol{#1}}
	\newcommand{\args}[1]{\ve{#1},\ve{\bar{#1}}}
	\newcommand{\mes}[1]{\mathcal{D}\hspace{-2pt}\left[\args{#1}\right]}
	\newcommand{\ii}{\iota}
	\newcommand{\mnm} {\ensuremath{\mathbb{M}}}
	\newcommand{\ncr}[2]{\begin{pmatrix}#1\\#2\end{pmatrix}}

	\DeclarePairedDelimiter\bra{\langle}{\rvert}
	\DeclarePairedDelimiter\ket{\lvert}{\rangle}


	\begin{abstract}
			We show that odd order R\'enyi entropies $S^{(2q+1)}$ of a system of interacting scalar fields can be calculated as the free energy of $2q+1$ replicas of the system with additional quadratic inter-replica couplings in the subsystem at the time of measurement of the entropy. These couplings replace boundary field matching conditions.  This formalism works both in and out of thermal equilibrium, for closed as well as open quantum systems, and provides a general dictionary between measurable correlation functions and entanglement entropy. $S^{(2q+1)}$ can be analytically continued to calculate the von Neumann entropy $S^{\mathrm{vN}}$. We provide an exact formula relating $S^{(2q+1)}$ and $S^{\mathrm{vN}}$ with correlation functions in a non-interacting theory. For interacting theories, we provide rules for constructing all possible Feynman diagrams for $S^{(2q+1)}$. We show that the boundary matching conditions cannot be completely eliminated while calculating R\'enyi entropies of even order due to presence of zero modes in replica space.
	\end{abstract}
	\title{Entanglement Entropy from Correlation Functions of Scalar Fields in and out of Equilibrium}
	\author{Mrinal Kanti Sarkar}\email{mrinal.sarkar@theory.tifr.res.in}
	\affiliation{Department of Theoretical Physics, Tata Institute of Fundamental
		Research, Mumbai 400005, India.}
	
	\author{Saranyo Moitra} 
	\affiliation{Department of Theoretical Physics, Tata Institute of Fundamental
		Research, Mumbai 400005, India.}
	
	\author{Rajdeep Sensarma} 
	\affiliation{Department of Theoretical Physics, Tata Institute of Fundamental
		Research, Mumbai 400005, India.}
	
	\date{\today}

	\maketitle
	\section{Introduction}
	
	Entanglement entropy of a many body system and its dynamics play an important role in the modern understanding of a wide range of physical phenomena: our basic understanding of quantum mechanics~\cite{Schrodinger_1935,Schrodinger_1936, EPR_main_paper, BohmEPR,Bells_inequality,CHSH,Aspect_1,Aspect_2,Zeilinger_expt}, classification of phases in quantum many body systems~\cite{LAFLORENCIE20161,Jiang2012, LevinWen_Topo,HastingsMelko,Wilczek, Calabrese_2009,CalabreseCardy,Casini_2009}, emergence of quantum gravity ~\cite{RyuTakayanagi2006}, thermalization~\cite{SrednickiETH, Deutsch_2018,LAFLORENCIE20161,Nandkishore,HusePal,abanin2019colloquium}, and development of quantum chaos~\cite{HuPRE70_2004, CasatiPRA71_2005, Franz_2_2020,CiracPRL126_2021} in quantum systems. Entanglement dynamics, especially in noisy open quantum systems, form a crucial aspect of understanding practical quantum computers~\cite{preskill2018quantum,gottesman1998heisenbergrepresentationquantumcomputers}. In a very limited number of cases, Rényi entanglement entropies of many-body systems have been measured experimentally~\cite{Islam2015_expt,Schreiber_MBL,Tajik2023}.

	Scalar field theory is a paradigmatic model for many Bosonic degrees of freedom in the continuum. They play a crucial role in high-energy physics~\cite{Srednicki_2007_book,Peskin_1995ev} as well as models of early dynamics of the universe~\cite{Bassett2006}. In condensed matter physics, they arise in systems where a continuous symmetry is broken~\cite{Sachdev_2011_Book}, e.g., solids (phonons~\cite{doniach}), magnets (magnons~\cite{magnon}), superfluids (phase fluctuations~\cite{superfluid_anderson,superfluid_book}), and so on. Scalar fields can also be used to model radiation in a cavity and resulting light-matter interactions~\cite{cvtqedexp1,cvtqedexp2,cvtqedrev1,cvtqedrev2}, leading to a wide range of dynamical phenomena. 
	
	The ubiquitous nature of scalar field theory has also made it a basic testing ground for developing quantum field theory based methods for calculating entanglement entropy. Early work ~\cite{Bombelli,Srednicki} on free scalar fields in thermal equilibrium calculated the scaling of entanglement entropy in free systems. In one spatial dimension, CFT based methods~\cite{Wilczek,Calabrese_2009,Casini_2009} showed logarithmic scaling of entanglement in critical systems in their ground state. More recently, it has been shown~\cite{Huerta_Velde_2023,Mrinal_Sl} that angular momentum resolved entanglement entropy of free scalar Bosons in $d>1$ can show interesting logarithmic scaling in the massless limit, which can be used to distinguish the critical state from the gapped states. 
	Entanglement in interacting $\phi^4$ theory has been studied in particular limits using different approaches: e.g., the large-$N$ limit near the critical point~\cite{SubirON,SubirWitczakKrempa}, perturbation theory for a particular entanglement cut~\cite{Hertzberg_2013}, orbifold methods~\cite{Iso_2021}, and variational techniques~\cite{COTLER_Variational}.
	 
	In spite of its seminal importance, the exact relation between correlation functions and entanglement entropy for a theory of interacting scalar Bosons is not yet known. Such a relation can lead to analytic and semi-analytic approximations for entanglement entropy of interacting scalar fields. It would also lead to more efficient approximate numerical evaluations of entanglement, since efficient algorithms to numerically calculate correlation functions already exist. Finally, the relation between correlations and entanglement is not just about theoretical understanding; it may be the only viable way to experimentally measure entanglement entropy. For free scalar fields, quantum state tomography in ultracold atomic systems has recently been used to measure correlations, and the known relation between correlators and entanglement has been used to infer the scaling of entanglement entropy of the system~\cite{Tajik2023}.
	
	In this paper, we formulate a new field-theoretic way to calculate the entanglement entropy of a system of interacting scalar fields in terms of the measurable correlation functions in the system. Our method works both in and out of thermal equilibrium. It provides a new method for calculating dynamics of entanglement in these systems, which can be useful for a wide range of problems, from dynamics across quantum phase transitions~\cite{RMP_polkovnikov} to Floquet~\cite{Moessner2017} dynamics to early evolution of the universe~\cite{Bassett2006}. 
	
	We (i) construct a local Hilbert space using an alternate quantization of scalar fields. This is the first step towards calculating the reduced density matrix and entanglement entropy. (ii) use a Wigner function based approach to show that R\'enyi entropies of odd order, $S^{(2q+1)}$, are given by the free energy of $2q+1$ replicas of the system, with inter-replica couplings between the fields $\phi$ and their conjugate momenta $\pi$ in the subsystem $A$ at the time of measurement of entanglement entropy (EE). {\it The inter-replica couplings are quadratic, and no additional field matching boundary conditions are required in this case.} This is much easier to handle than the boundary conditions in standard field theoretic formulation~\cite{CalabreseCardy}, and has a wider applicability. (iii) show that one cannot get rid of all field matching conditions while calculating even order R\'enyi entropies. This is due to a particular zero mode in the replica space. However, we manage to replace the boundary conditions for each replica by two boundary conditions involving all the replica fields and their conjugate momenta. (iv) show that the odd R\'enyi entropies can be analytically continued to obtain the von Neumann entanglement entropy $S^{\mathrm{vN}}$ of the system. (v) derive an exact relation between EE and measurable correlation functions for non-equilibrium dynamics of Gaussian open quantum systems. (vi) provide the full set of Feynman diagrams to calculate EE of an interacting system and the Feynman rules that go along with them. These are constructed explicitly in terms of the standard Schwinger-Keldysh correlators of the theory in absence of any replica, and the interaction vertices. We provide explicit expressions for perturbative corrections to $S^{(n)}$ in a $\phi^4$ theory. Our formalism works for arbitrary interacting systems (open or closed) in and out of thermal equilibrium and provides a detailed dictionary between measurable correlation functions which occur in field theories with a single replica and entanglement entropy. We expect that this dictionary will become a template for measuring EE indirectly through measurements of correlation functions. 

	This paper is organized as follows: In Section~\ref{Coherent:States}, an alternate quantization of scalar fields is developed, which allows the construction of a local Hilbert space in terms of the local fields and their conjugate momenta. In Section~\ref{WCF:Sn},  we define Wigner characteristic functions (WCF) and show how the entanglement entropy can be written in terms of integrals over these functions. In Section~\ref{Sn:FieldTheory}, we express the WCF as a Schwinger-Keldysh partition function in presence of sources and hence construct a field theory for entanglement entropy. We show that the sources can be explicitly integrated out and R\'enyi entanglement entropy is the free energy of replicas with quadratic inter-replica couplings. In Section~\ref{Gaussian}, we apply our formalism to a Gaussian theory and find an explicit relation between Green's functions and entanglement entropy.  In Section~\ref{Interactions}, we construct Feynman diagrams for calculating entanglement entropy in an interacting system. Finally, we conclude by summarizing our results in Section~\ref{Conclusions}.

\section{ \label{Coherent:States} Scalar field theory and local coherent states}
        
We consider an interacting theory of scalar fields in $d$ spatial dimensions defined by the action,
\beq
\mathcal {S}= \frac{1}{2}\int dt~ \int d^d \rr~ \phi(\rr,t) \left[-\partial_t^2+\nabla^2-m^2\right] \phi(\rr,t) + \mathcal {S}_{\mathrm{int}}
\eeq
where $m$ is the bare mass and $\mathcal {S}_{\mathrm{int}}$ represents the interaction between the fields. A typical form of $ \mathcal {S}_{\mathrm{int}}= -\frac{\lambda}{4!} \int dt \int d^d \rr~\phi^4(\rr,t)$ defines the well known $\phi^4$ theory \cite{Srednicki_2007_book}. However, our formulation does not depend on the particular form of $\mathcal S_{\mathrm{int}}$; e.g., it can accommodate more complicated interactions which can lead to dipole conservation in the system \cite{Pretko_dipole,Mursalin_dipole}. We will supplement this action with a momentum cutoff $\Lambda$ for UV regularization. In standard field theory, one considers the fields in momentum space $\phi(\kk,t)$ and its conjugate field $\pi(\kk,t)=\dot{\phi}(\kk,t)$, and expands them in terms of the creation and annihilation operators,
\bqa
\phi(\kk,t)&=&\frac{1}{\sqrt{2\omega_\kk}} \left[e^{-i\omega_\kk t} a_\kk +e^{i\omega_\kk t} a^\dagger_{-\kk}\right],\\
\no \pi(\kk,t)&=&-i\sqrt{\frac{\omega_\kk}{2}} \left[e^{-i\omega_\kk t} a_\kk -e^{i\omega_\kk t} a^\dagger_{-\kk}\right],
\eqa
where $\omega_\kk=\sqrt{k^2+m^2}$ is the dispersion of the non-interacting modes, and $[a_\kk,a^\dagger_{\kk'}]=\delta^{(d)}(\kk-\kk')$. The number states/coherent states corresponding to $a_\kk$ provide a convenient basis to construct a Hilbert space in this case.

A key requirement for computing the entanglement entropy is the construction of a local Hilbert space that factorizes as a tensor product of Hilbert spaces associated with the subsystem $A$ and its complement $A^c$. One might naively guess that $a_\rr=\int\frac{d^d\kk}{(2\pi)^d} a_\kk e^{i\kk.\rr}$ can be used to construct such a Hilbert space. However, since $a_\kk = \frac{e^{i\omega_\kk t}}{\sqrt{2}}\left[\sqrt{\omega_\kk} \phi(\kk,t)+ i \frac{\pi(\kk,t)}{\sqrt{\omega_\kk}}\right]$, it is clear that the relation between $a_\rr,a^\dagger_\rr$ and $\phi(\rr,t),\pi(\rr,t)$ would be non-local in space with kernels given by Fourier transforms of $\omega_\kk^{\pm \frac{1}{2}}$. The extent of non-locality will increase with decreasing $m$ (increasing correlation length), and the kernels will have power law decays in the massless/critical limit. Thus, this construction would necessarily require knowledge of correlations of $\phi$ and $\pi$ in $A^c$ to construct the local Hilbert space in $A$. Hence, this Hilbert space is not useful for the calculation of entanglement entropy in terms of the correlations of the fields.

To circumvent the issue of non-locality, we define creation/annihilation operators $b^\dagger_\rr$, $b_\rr$ as
\begin{equation}
	\begin{split}
	b_\rr &=\frac{1}{\sqrt{2}}\left[ \sqrt{E_0} \phi(\rr,t_0)+i \frac{\pi(\rr,t_0)}{\sqrt{E_0}}\right],\\
	b^\dagger_\rr &=\frac{1}{\sqrt{2}}\left[ \sqrt{E_0} \phi(\rr,t_0)-i \frac{\pi(\rr,t_0)}{\sqrt{E_0}}\right],
	\end{split}\label{bbdaggereqn}
\end{equation}
where $[b_\rr,b^\dagger_{\rr'}]=\delta^{(d)}(\rr-\rr')$. Here, $t_0$ is the time at which the EE is measured, and $E_0$ is an arbitrary energy scale that we have introduced to obtain dimensionless $b, b^\dagger$. We can take $E_0$ to be the largest energy scale in the problem. In the end, we will show that the scale $E_0$ drops out when the EE is expressed in terms of correlators of $\phi(\rr,t)$ and $\pi(\rr,t)$, but plays an important role in conceptualizing a local Hilbert space and formulating the calculation of EE. The operator algebra of $b_\rr$ allows us to construct number states and coherent states, which are related to the local fields. The non-interacting Hamiltonian is not diagonal in this basis; however, in this case, it is more important to have a {\it local} basis than a basis that diagonalizes the non-interacting Hamiltonian.

It is instructive to compare the case of scalar fields with Schr\"odinger Bosons, which are represented by complex fields with an action of the form $\mathcal S= \int d^d\rr~ \int dt~ [\phi^\ast(\rr,t) i\partial_t \phi(\rr,t) -H(\phi^\ast,\phi)]$. For Schr\"odinger Bosons, the conjugate field to $\phi(\rr,t)$ is $\phi^\ast(\rr,t)$. Since the field and its conjugate have the same dimension (contrast this with scalar fields where $\phi$ and $\pi=\dot{\phi}$ do not have the same dimension), one can add them to construct creation/annihilation operators without introducing any scale (independent of the Hamiltonian). The extra scale for the scalar fields is thus a direct consequence of the second-order time derivatives in the equation of motion. We would like to note that the $O(N)$ fields can be thought of as $N$ real scalar fields, and the generalization of this construction to $O(N)$ theories is straightforward. In the next section, we will use Wigner Functions of density matrices defined in this local Hilbert space to calculate R\'enyi entanglement entropy of scalar fields.
\section{\label{WCF:Sn}Rényi entropy from Wigner Characteristic function}

The $n^{\mathrm{th}}$ order R\'enyi entanglement entropy is given by 
\begin{equation}
	S^{(n)}=\frac{1}{1-n}\ln~(\operatorname{Tr}\left[  \hat{\rho}_A^n\right] ),
\end{equation}
 where $\hat{\rho}_A=\operatorname{Tr}_{A^c}[\hat{\rho}]$ denotes the reduced density matrix (RDM) of the subsystem $A$, obtained by tracing out its complement $A^c$ from the full system $A \cup A^c$ with density matrix $\hat{\rho}$. The von Neumann entropy is obtained by analytic continuation,
 \begin{equation}
 			S^{\mathrm{vN}}=\lim_{n\to1}S^{(n)}=-\operatorname{Tr}[\hat{\rho}_A \ln~\hat{\rho}_A ].
 \end{equation}

To calculate $S^{(n)}$, we first use the local operators ($b_\rr, b^\dagger_\rr$) to define  a displacement operator for subsystem $A$,
\begin{equation}
	\hat{\textrm D}_A(\xib)=e^{\int_A d^d\rr~\left[\xi_\rr b^\dagger_\rr-\xi^\ast_\rr b_\rr\right]},
\end{equation} where the integral in the exponent is restricted to the subsystem $A$. It was shown by Cahill and Glauber~\cite{Cahill_Glauber_1969} that any bounded bosonic operator $\hat{F}$ with support only in the Hilbert space of $A$ can be expanded in terms of displacement operators as:
\begin{equation}
	\hat{F}=\int \mathscr{D}[\xib]~\chi_F(\xib)~\hat{\textrm D}_A(-\xib).
\end{equation}
Here, $\chi_F(\xib)=\operatorname{Tr}[\hat{F}~\hat{\textrm D}_A(\xib)]$ is known as the characteristic Weyl symbol of $\hat{F}$ and $ \mathscr D[\xib]=\prod_{\rr\in A}\frac{d\xi_\rr d\xi^\ast_\rr}{2\pi}$.  For the reduced density matrix $\hat{\rho}_A$, the corresponding  function is called the Wigner Characteristic Function (WCF) 
\begin{equation}
	\chi_A(\xib,t_0)=\operatorname{Tr}_A[\hat{\rho}_A(t_0)\hat{\textrm D}_A(\xib)]=\operatorname{Tr}[\hat{\rho}(t_0)\hat{\textrm D}_A(\xib)].
\end{equation} 
Thus, the WCF is the expectation of $\hat{\textrm D}_A$ in the full density matrix~$\hat{\rho}$~\cite{AhanaPRL,AhanaPRA, moitra2020entanglement,Saranyo_Building_Entanglement}.

We can use this to expand $\hat{\rho}_A^n$ in powers of $\chi_A$ and $\hat{\mathrm D}_A$. Finally, we employ the combination rule for displacement operators $\hat{\textrm D}_A(\xib)\hat{\textrm D}_A(\etab)=\hat{\textrm D}_A(\xib+\etab)e^{\frac{1}{2}\int_A d^d\rr~\left[\xi_\rr\eta_\rr^\ast-\xi_\rr^\ast\eta_\rr\right]}$, together with trace identity $\operatorname{Tr}[\hat{\textrm D}_A(\xib)\hat{\mathrm D}_A(\etab)]=\delta(\frac{\xib+\etab}{\pi})$, to write 
\begin{widetext}
	\bqa
	e^{(1-n)S^{(n)}}&=&\left[ \int\prod_{\alpha=1}^{n-1}\mathscr D[\xib^{(\alpha)}] \right ] \chi_A(\xib^{(1)}) \chi_A(\xib^{(2)})... \chi_A(\xib^{(n-1)})\chi_A(- \xib^{(1)}-\xib^{(2)}...~ -\xib^{(n-1)}) ~\times~ e^{\frac{1}{2} \mathbf{\zeta}^\dagger \mathbb{K}_{n-1}\otimes \hat{P}_A \mathbf {\zeta}}.
	\label{eqsnbeta}
	\eqa
\end{widetext}
Here, $\alpha$ denotes replica index, ${\zeta}^\dagger = \left[\xib^{*(1)}, \xib^{*(2)}... ~\xib^{*(n-1)}\right]$ and $\mathbb{K}_{n-1}$ is an $(n-1)$ dimensional real skew-symmetric skew-circulant matrix in replica space,
\beq
\mathbb{K}_{n-1}=\left( \begin{array}{ccccc}
	0 & -1 & -1& ... & ...\\
	1 &0 &  -1 & -1& ...\\
	1 &1 & 0 & -1& ...\\
	... & 1 & 1& 0& ...\\
	... & ... & ...& ...& ...
\end{array}\right).
\eeq 
The operator $\hat{P}_A$ is the projection onto the subsystem $A$. As a special case, it is easy to see that the second R\'enyi entropy $S^{(2)}$ can be expressed as
\begin{equation}
	\begin{split}
		e^{-S^{(2)}}&=\operatorname{Tr}[\hat{\rho}_A^2]=\int\mathscr{D}[\xib]~\chi_A(\xib)\chi_A(-\xib).\\
	\end{split}
\end{equation}
Eq.~\eqref{eqsnbeta} is the key result which relates the Wigner characteristic functions to the R\'enyi entanglement entropy of scalar Bosons. It involves a product of $n$ WCFs with different arguments, a Gaussian factor, and integrals over all arguments. Our next task is to calculate $\chi$ within a field-theoretic approach, which will lead to a field-theoretic formulation of $S^{(n)}$. We will discuss this in the next Section.

\section{\label{Sn:FieldTheory}Rényi entropy, Wigner Characteristic function, and Keldysh partition function}

In this section, we want to develop a general formalism that can tackle the non-equilibrium dynamics of entanglement entropy. A natural choice is to work with a non-equilibrium Schwinger-Keldysh (SK) field theory. We will map the WCF of the RDM to a Keldysh partition function of the system in presence of linear sources. The sources will be related to the arguments of the WCF. The integrals in Eq.~\eqref{eqsnbeta} will then lead to a field-theoretic calculation of the R\'enyi entropies and, by analytic continuation, a calculation of the von Neumann entropy of a subsystem in a real scalar field theory. 

\begin{figure}[t] \includegraphics[width=\columnwidth]{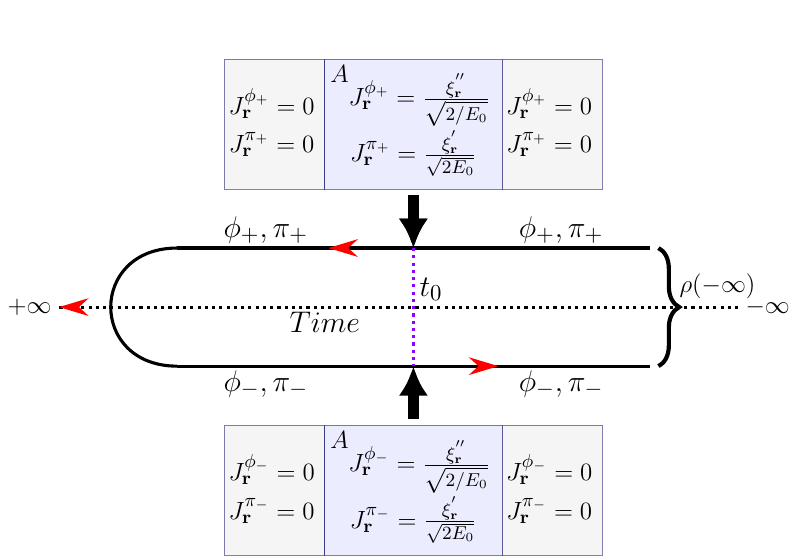}
	\caption{Schematic representation of the WCF calculation on Keldysh time contour. It consists of a forward time evolution ($``+"$ contour) extending from $-\infty$ to $+\infty$ and a backward time evolution ($``-"$ contour) returning from $+\infty$ to $-\infty$. Fields defined on the forward and backward branches are denoted as $\phi_+ $ and $\phi_-$, respectively. At the measurement time $t_0$, sources $J_\rr^{\phi_+}$ and $J_\rr^{\pi_+}$ are coupled to the fields $\phi_+$ and $\pi_+$ respectively only within subsystem $A$ on the forward contour, while sources $J_\rr^{\phi_-}$ and $J_\rr^{\pi_-}$ are coupled to the fields $\phi_-$ and $\pi_-$ only within subsystem A on the backward contour. The corresponding partition function gives the WCF of the RDM.}
	\label{Keldysh_contour_and_source_insertion}
\end{figure}

In an SK field theory, the forward time evolution of the kets and the backward time evolution of the bras are combined to construct the time evolution of the density matrix. This leads to a field theory with two fields $(\phi_\pm)$ at each spacetime point, where $``+"$ and $``-"$ indicate the time contours for forward and backward time evolutions (see the $``\pm"$ time contours in  Fig.~\ref{Keldysh_contour_and_source_insertion} for a schematic visualization).

To make further progress, we note that in Ref.~\onlinecite{AhanaPRL} and~\onlinecite{AhanaPRA}, it was shown that for Bosons obeying a Schrodinger equation,  $\chi_A(\xib)$ can be evaluated as a SK  partition function in presence of sources. The insertion of the displacement operator is equivalent to coupling sources $ \mp i \xi_\rr/2$ to the corresponding complex field $\psi^\ast_{\pm}(\rr,t_0)$ and the source $\pm i \xi^\ast_\rr/2$ to $\psi_{\pm}(\rr,t_0)$ on the forward (backward) contours. The sources are turned on only in the subsystem $A$ at $t=t_0$, when the Wigner characteristic is measured. 

The extension of this construction to the local coherent states of scalar fields is straightforward: we couple sources $\mp i \xi_\rr/2$ to fields corresponding to $b^\dagger_\rr$ and $\pm i \xi^\ast_\rr/2$ to fields corresponding to $b_\rr$ on the $``\pm"$ time contours for $\rr\in A$ at $t=t_0$. This gives the Wigner characteristic function for the reduced density matrix of the scalar fields. Using the relation between $b_\rr$ and the scalar fields defined in Eq.~\eqref{bbdaggereqn}, the source terms in the action have the form 
\begin{widetext}
\bqa\label{source_incertion_equn}
\no i \delta \mathcal S &=&\int_A d^d\rr ~  \frac{\xi_\rr}{2\sqrt{2}}\left[ \sqrt{E_0} [\phi_+(\rr,t_0)+\phi_-(\rr,t_0)]-i \frac{\pi_+(\rr,t_0)+\pi_-(\rr,t_0)}{\sqrt{E_0}}\right]\\
 & -& \frac{\xi^\ast_\rr}{2\sqrt{2}}\left[ \sqrt{E_0} [\phi_+(\rr,t_0)+\phi_-(\rr,t_0)]+i \frac{\pi_+(\rr,t_0)+\pi_-(\rr,t_0)}{\sqrt{E_0}}\right] \\
\no &=&\int_A d^d\rr ~ \frac{(\xi_\rr-\xi^\ast_\rr)}{2\sqrt{2}} \sqrt{E_0} [\phi_+(\rr,t_0)+\phi_-(\rr,t_0)]-i\frac{(\xi_\rr+\xi^\ast_\rr)}{2\sqrt{2}} \frac{\left[\pi_+(\rr,t_0)+\pi_-(\rr,t_0)\right]}{\sqrt{E_0}}.
\eqa
\end{widetext}

The first thing to note in Eq.~\eqref{source_incertion_equn} is that the sources are coupled only to the symmetric combination of fields taken from the $``+"$ and $``-"$ contours. It is then easier to work with the symmetric (or ``classical'') combination ($\phi_{s} = \frac{1}{\sqrt{2}}[\phi_++\phi_-]$, $\pi_{s} = \frac{1}{\sqrt{2}}[\pi_++\pi_-]$ ) and the antisymmetric  (or ``quantum'') combination  ( $\phi_{a} = \frac{1}{\sqrt{2}}[\phi_+-\phi_-]$, $\pi_{a} = \frac{1}{\sqrt{2}}[\pi_+-\pi_-]$ ) of the fields. It is also useful to decompose the complex sources into real and imaginary parts $\xi_\rr = \xi^{'}_\rr +i \xi^{''}_\rr$ to rewrite the source terms as
\beq
	\delta \mathcal S=~\int_A d^d\rr ~ \left[\xi^{''}_\rr\sqrt{E_0} \phi_s(\rr,t_0)-\frac{\xi^{'}_\rr}{\sqrt{E_0}} \pi_s(\rr,t_0)\right].
\eeq

The WCF of the RDM is the Keldysh partition of the theory in presence of sources coupled to the symmetric fields and their conjugate momenta in the subsystem $A$ at the time of measurement $t_0$. The real part of the argument couples to $\pi_s/\sqrt{E_0}=\dot{\phi}_s/\sqrt{E_0}$, while its imaginary part couples to $\sqrt{E_0}\phi_s$, giving
\beq
\chi_A(\xib) = \int \mathcal D[\phi_{s,a}] e^{i \left(\mathcal S[\phi_{s,a}] +\delta \mathcal S[\phi_s,\pi_s]\right)},
\eeq
where $\mathcal S$ is the original Keldysh action of a possibly interacting theory.

The scale $E_0$ appears explicitly in the formula of the Wigner characteristic, as it is inherent in the definition of the displacement operator we used. We will see that this scale will drop out when we consider the integrals over $\xi$ needed to calculate entanglement entropy, as shown in Eq.~\eqref{eqsnbeta}. To see this, note that if we define $J^\pi_\rr= \frac{\xi^{'}_\rr}{\sqrt{E_0}}$ and $J^\phi_\rr=\xi^{''}_r\sqrt{E_0} $, the integral measure in Eq.~\eqref{eqsnbeta} can be written as $ ~ d\xi_\rr d\xi_\rr^\ast/2 = d\xi^{'}_\rr~ d \xi^{''}_\rr =d J^{\phi}_\rr~ d J^{\pi}_\rr$. 
Switching to the source fields $J^{\phi,\pi}$, we rewrite
\beq
 	\chi_A(\JJ^\phi,\JJ^\pi) = \int {\cal D}[\phi_{s,a}]e^{i \mathcal S[\phi_{s,a}] +i\int_A d^d\rr ~ \left[J^\phi_r \phi_s(\rr,t_0)-J^\pi_r \pi_s(\rr,t_0)\right],}
 	\label{eqchiwJ}
 \eeq 
where $\mathcal{D}[\phi_{s,a}]=\mathcal{D}[\phi_{s}]\mathcal{D} [\phi_{a}]$, with $\mathcal{D}$ being the usual field theoretic measure. Furthermore, defining ${\cal J}^T=\left(\JJ^{\phi(1)},\JJ^{\pi(1)},..\JJ^{\phi(n-1)},\JJ^{\pi(n -1)}\right)$, the exponential factor in Eq.~\eqref{eqsnbeta} can be written as
\beq
\nonumber \frac{i}{2}{\cal J}^T\mathbb{K}_{n-1}\otimes \left[\begin{array}{cc} 0 &-\hat{P}_A\\
	\hat{P}_A & 0 \end{array}\right]{\cal J}.
\eeq

Putting all this together in Eq.~\eqref{eqsnbeta}, we then get
\begin{widetext}
\begin{equation}
\begin{split}\label{eqSnJ}
 e^{(1-n)S^{(n)}}&= \int \prod_{\alpha=1}^n {\cal D}[\phi^{(\alpha)}_{s,a}]~ \exp\left[i\sum_{\alpha=1}^n \mathcal{S}[\phi^{(\alpha)}_{s,a}]\right]\\
 &\times\int \prod_{\alpha=1}^{n-1} ~{\mathscr D}[{\bf J^{(\alpha)}}]~\exp\left[i\sum_{\alpha=1}^{n-1} \int_A d^d\rr ~J_\rr^{\phi(\alpha)} (\phi^{(\alpha)}_s(\rr,t_0)-\phi^{(n)}_s(\rr,t_0))-J_\rr^{\pi(\alpha)} (\pi^{(\alpha)}_s(\rr,t_0)-\pi^{(n)}_s(\rr,t_0))\right]\\
 &\times~ \exp\left[\frac{i}{2}{\cal J}^T\mathbb{K}_{n-1}\otimes \left[\begin{array}{cc} 0 &-\hat{P}_A\\
 	\hat{P}_A & 0 \end{array}\right]{\cal J}\right].\\
\end{split}
\end{equation}
\end{widetext}
Here ${\mathscr D}[{\bf J^{(\alpha)}}]= \prod_{\rr\in A} \frac{dJ^{\phi(\alpha)}_\rr~d J^{\pi(\alpha)}_\rr}{\pi}$.
 The Eq.~\eqref{eqSnJ} provides the exact starting point from which one can proceed to construct a diagrammatic representation of the entanglement entropy of a scalar field theory in or out of equilibrium. The final expression for $S^{(n)}$ involves integrals over the classical source fields $\JJ$ as well as over the quantum fields $\phi$. Note that there are $n$ copies of $\phi$ but $n-1$ $\JJ$s. Eq.~\eqref{eqSnJ} doesn't depend on the detailed form of the action $\mathcal{S}$, hence it holds for any generic non-equilibrium system. 
 
 The Gaussian integrals over the sources require the inverse of the matrix $\mathbb{K}_{n-1}$. Since $\mathbb{K}_{n-1}$ is an $(n-1)$ dimensional real antisymmetric matrix, its eigenvalues are purely imaginary and occur in complex conjugate pairs. For odd $(n-1)$, i.e., for even $n$, $\mathbb{K}_{n-1}$ would necessarily have a zero eigenvalue and would not be invertible. Hence, we will treat the case of even and odd order R\'enyi entropies separately. We first focus on odd $n$, where $\mathbb{K}_{n-1}$ is invertible. In this case, the Gaussian integrals over the source fields can be performed exactly (see Appendix~\Ref{A_odd_Renyi} for details) to give
\begin{equation}
\begin{split}
	\frac{e^{(1-n)S^{(n)}}}{2^{(n-1)V_A}}=\int\prod_{\alpha=1}^{n} \mathcal D[\phi^{(\alpha)}_{s,a}]~e^{i\sum\limits_{\alpha=1}^{n}\mathcal{S}[\phi_{s,a}^{(\alpha)}]~+i\mathcal{S}_{ent}[\phi_{s},\pi_s]},\label{EqSent}\\
\end{split}
\end{equation}
where $\mathcal{S}[\phi_{s,a}^{(\alpha)}]$ denotes the contribution from the action of $\alpha^{\mathrm{th}}$ replica, $V_A$ is the volume of the subsystem $A$, and the entangling action $S_{ent}$ is given by
\begin{widetext}
\begin{equation}
	\label{eq:Sent:odd}
	\mathcal{S}_{ent}(t_0)=\frac{1}{2}\sum_{\alpha,\beta=1}^{n}\int_Ad^d\rr~\left[\begin{array}{c}{\bf \phi}^{(\alpha)}_s(\rr,t_0),{\bf \pi}^{(\alpha)}_s(\rr,t_0)\\ \end{array}\right]\mathbb{L}^{\alpha\beta}
	\left[\begin{array}{c}{\bf \pi}^{(\beta)}_s(\rr,t_0)\\ -{\bf \phi}^{(\beta)}_s(\rr,t_0)\\ \end{array}\right].
\end{equation}
\end{widetext}
Here, the coupling matrix $\mathbb L$ has a skew-symmetric circulant structure $\mathbb{L}^{\alpha\beta} = (-1)^{\alpha-\beta}\left[\Theta(\alpha-\beta)-\Theta(\beta-\alpha)\right]$ i.e. in the replica space,
\begin{equation}
	\mathbb L= 
	\begin{bmatrix}
		\ \: \: \: 0 & \:\:\: 1 & -1 &\:\:\:1 &... &-1\\
		\ -1&  \: \: \:0 & \:\:\:1 &-1&... & \: \: \: 1\\
		\ \:\:\:1& -1  & \: \: \:0 &\:\:\:1&... &-1\\
		-1& \:\:\:1  & -1 &\: \: \:0&... &\: \: \: 1\\
		\ \: \: \: . & \: \: \: . & \: \: \: .&\: \: \: .& &  \: \: \: . \\
		
		\ \: \: \:1& -1 &\: \: \:1 &-1& ...& \: \: \:0 \\
	\end{bmatrix}_{n\times n}.
\end{equation}
Thus, the odd order R\'enyi entropy of scalar fields is the free energy of $n$ independent Keldysh replicas coupled by $\mathcal{S}_{ent}$. Note that
\begin{itemize}
\item the arbitrary scale $E_0$ has dropped out of the description of entanglement entropy (Eq.~\eqref{EqSent} and \eqref{eq:Sent:odd}).

\item $\mathcal{S}_{ent}$ couples the fields in one replica with the conjugate momenta in the other replica.

\item $\mathcal{S}_{ent}$ is purely real and does not couple fields or their conjugate momenta in the same replica.

\item $\mathcal{S}_{ent}$ only couples fields and momenta in the subsystem $A$. The coupling is local in space.

\item the coupling $\mathcal{S}_{ent}$ exists only at time of measurement $(t_0)$ of the entanglement entropy. Hence, this cannot be gauged away as a boundary term.

\item $\mathcal{S}_{ent}$ couples symmetric Keldysh fields with symmetric versions of conjugate momenta. Such terms are forbidden in usual Schwinger-Keldysh field theories. 
\end{itemize}

At this point, it is useful to note that the analytic continuation of the R\'enyi entropies to von Neumann entropy can be done solely in terms of the odd R\'enyi entropies, i.e.,
\beq
S^{\mathrm{vN}} = \lim_{q \rightarrow 0} S^{(2q+1)}.
\eeq
So, if one is interested in calculating the von Neumann entanglement entropy, one can avoid the complications associated with evaluating even-order R\'enyi entropies. It is also easy to see that one can add a flavour index to the fields, and the generalization to $O(N)$ theory is straightforward. 
\begin{figure}[t] \includegraphics[width=0.8\columnwidth]{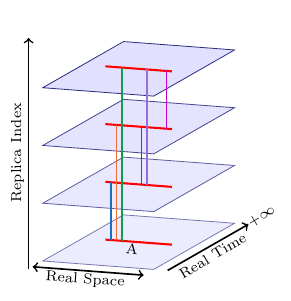}
	\caption{Schematic representation of inter-replica coupling in $\mathcal{S}_{ent}$ between symmetric component of the fields $(\phi_s)$ in one replica with the conjugate fields $(\pi_s)$ in other replica only within the subsystem $A$ at the time $t_0$. Note that this is an all-to-all coupling with strength $\pm 1$. There is no self-replica coupling.}
	\label{Replica_coupling}
\end{figure} 

The calculation of even order ($n=2q$) R\'enyi entropy is more subtle. A straightforward evaluation of the Gaussian integral over ${\mathcal J}$ is not possible due to the singular nature of $\mathbb{K}_{2q-1}$. In this case, we treat the zero mode of $\mathbb{K}_{2q-1}$ separately, which leads to the following boundary conditions in $\phi_s$ and $\pi_s$ in the subsystem $A$ at time $t_0$:
\begin{equation}
	\begin{split}
	\sum_{\alpha=1}^{n}(-1)^{\alpha}\phi_s^{(\alpha)}(\rr,t_0)&=0 ~~~\forall \rr \in A\\
		\sum_{\alpha=1}^{n}(-1)^{\alpha}\pi_s^{(\alpha)}(\rr,t_0)&=0~~~\forall \rr \in A
	\end{split}
\end{equation}
i.e., the sum of the fields over even replicas must equal the sum over odd replicas, and an analogous constraint applies to the conjugate momenta.
 
The standard field-theoretic calculations of entanglement entropy require field matching between successive replicas. In our formalism, this is replaced by a single boundary condition for even R\'enyi entropies $S^{(2q)}$ while no field matching is required for odd R\'enyi entropies $S^{(2q+1)}$. One can then do the usual Gaussian integrals over the non-singular eigenmodes of $\mathbb{K}_{2q-1}$. The final result for $n=2q$ can be written as (see Appendix~\Ref{A_even_Renyi} for details),
\begin{widetext}
	\begin{equation}
		\label{Sneven}
	\begin{split}
		\frac{e^{(1-n)S^{(n)}}}{2^{(n-1)V_A}\pi^{V_A}}&=\int\prod_{\alpha=1}^{n} \mathcal D[\phi^{(\alpha)}_{s,a}]~\exp\left[i\sum\limits_{\alpha=1}^{n}\mathcal{S}[\phi_{s,a}^{(\alpha)}]+\frac{i}{2}\sum\limits_{\alpha,\beta=1}^{n}\int_Ad^d\rr~\left[\begin{array}{c}{\bf \phi}^{(\alpha)}_s(\rr,t_0),{\bf \pi}^{(\alpha)}_s(\rr,t_0)\\ \end{array}\right]\mathbb{P}^{\alpha\beta}
		\left[\begin{array}{c}{\bf \pi}^{(\beta)}_s(\rr,t_0)\\ -{\bf \phi}^{(\beta)}_s(\rr,t_0)\\ \end{array}\right]\right]\\
		&\times	\prod_{\rr\in A}\delta\left[\sum_{\alpha=1}^{n}(-1)^{\alpha}\phi_s^{(\alpha)}(\rr,t_0)\right]\delta\left[\sum_{\alpha=1}^{n}(-1)^{\alpha}\pi_s^{(\alpha)}(\rr,t_0)\right]\\
	\end{split}
	\end{equation}
{\small
\begin{equation*}
	\text{where,}~\mathbb{P}^{\alpha\beta} =
	\begin{cases}
		(-1)^{\alpha-\beta}\left[\operatorname{sgn}(\alpha-\beta) - \dfrac{2(\alpha-\beta)}{\,n-1}\right], & 1 \leq \alpha,\beta \leq n-1, \\[1em]
		\frac{(-1)^{-\beta}}{n-1}\left(2\beta - n\right), & \alpha = n,\; 1 \leq \beta \leq n-1, \\[1em]
		-\frac{(-1)^{\alpha}}{n-1}\left(2\alpha - n\right), & 1 \leq \alpha \leq n-1,\; \beta = n,\\[1em]
		0, & \alpha = \beta = n.		
	\end{cases}
\end{equation*}
}
\end{widetext}

Here, the first $(n-1)\times(n-1)$ block of $\mathbb{P}$ is the pseudo inverse of $\mathbb{K}_{n-1}$. Thus, we have obtained expressions for the odd and even R\'enyi entropies.
In the next section, we will discuss entanglement of free scalar fields and show that we recover well-known results \cite{Srednicki,Bombelli,Casini_2009,Cotler2016}. This will act as a non-trivial check on our formalism.
\begin{figure*}[t] \includegraphics[width=\textwidth]{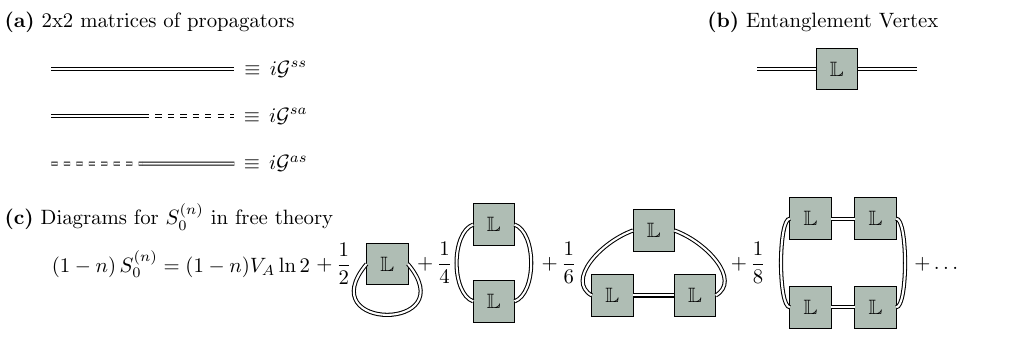}
	\caption{
		Feynman diagrams for evaluating $S^{(n)}_0$ in a free scalar field theory: Three non-zero $2 \times 2$ matrices of correlators $(\mathcal{G}^{pq};\,p,q\in\{s, a\})$, defined in the text, are represented by double lines in (a). Among these, $\mathcal{G}^{ss}$ is shown as fully solid double lines, $\mathcal{G}^{sa}$ as half-solid–half-dashed double lines, and $\mathcal{G}^{as}$ as half-dashed–half-solid double lines. The solid side corresponds to symmetric~(s) end, while the dashed side corresponds to antisymmetric~(a) end. (b) Diagrammatic representation of the entanglement vertex $S_{ent}$ which couples across replicas; both ends represent symmetric components of fields and their conjugate momenta inside the subsystem $A$ at time $t_0$. (c) R\'enyi entropies are expressed as the sum of connected ring diagrams containing various powers of $\mathbb{L}$ with the symmetry factor of each diagram indicated. Note that all propagators in (c) are $2 \times 2$ equal time Keldysh propagators $\mathcal{G}^{ss}(\rr,t_0;\rr',t_0)$ evaluated inside subsystem $A$ at $t_0$.}
	\label{Free_theory_diagrams}
\end{figure*}
\section{\label{Gaussian}Entanglement Entropy of Free Scalar Fields}

We first test our new formulae for the case of a theory of free scalar fields. An exact formula for the entanglement entropy of free scalar fields in their ground/thermal state is known in terms of correlation functions~\cite{Srednicki, Bombelli, Casini_2009}. This leads to a volume law scaling of $S^{(n)}$ and $S^{\mathrm{vN}}$  in thermal states at $T\neq 0$ and an area law scaling for ground states in $d>1$. For ground states in $d=1$, the same formula recovers the well-known logarithmic scaling of a 1+1 D CFT with coefficients given by the central charge of the theory~\cite{HOLZHEY1994443,CalabreseCardy}. 

Non-equilibrium dynamics of scalar fields is used to describe a wide range of problems, e.g., early expansion of the universe~\cite{Bassett2006}, phonons coupled to electronic baths~\cite{noneqscalar_anisimov,Mursalin2022}, dynamics of order parameters across phase transitions~\cite{LagunaPabloZurek,Halparin}, and photons in cavities~\cite{Weiher,cvtqwed_sqzlit3}. In this section, we will derive a formula that applies to a system of free scalar fields undergoing arbitrary (possibly non-Markovian) non-equilibrium dynamics (including an open quantum system coupled to external baths). We will see that certain correlation functions which vanish in an equilibrium theory are non-zero in a non-equilibrium situation. Hence, the general formula we will derive here will look different from the equilibrium formula for the entanglement entropy. The dynamics of entanglement in closed systems of scalar fields have been studied after a quench of parameters~\cite{Alba_calabrese, Cotler2016}; however, our formalism is applicable to open quantum systems with Markovian and non-Markovian baths~\cite{Power_law_Ahana,AhanaPRL}.

The Schwinger Keldysh action for a Gaussian theory is given by 
\begin{widetext}
	\beq
	\label{freescalar}
	\mathcal {S}_0= \frac{1}{2}\int d^d \rr ~\int d^d \rr' ~\int dt ~\int dt~[\phi_s(\rr,t),\phi_a(\rr,t)] \left[\begin{array}{cc}
		0 & G_0^{A-1}(\rr,t;\rr',t')\\
		G_0^{R-1}(\rr,t;\rr',t') & -\Sigma_0^K(\rr,t;\rr',t') \end{array}\right]\left[\begin{array}{c}
		\phi_s(\rr,t)\\
		\phi_a(\rr,t)\end{array}\right],
	\eeq
\end{widetext}
 where the general structure of the inverse retarded Green's function is of the form $G^{R-1}(\rr,t;\rr',t') =\left[-\partial_t^2+\nabla^2-m^2\right] \delta(t-t') \delta(\rr-\rr') - \Sigma^R(\rr,t;\rr',t')$. Here $m$ is the mass, and the retarded self-energy $\Sigma^R$ provides a non-local kernel whose real part leads to dressed coherence, while its imaginary part controls dissipation in the system. Additionally, one may need to consider an ultraviolet momentum cutoff $\Lambda$ to make the entanglement entropy well defined for the system. The Keldysh self-energy $\Sigma^K$ has contributions from both initial conditions and any noise generated by an external bath. Note that we are not assuming time-translation invariance in the system. 
 
 This Gaussian action leads  to the single-particle Green's functions 
 \begin{equation}
 	\begin{split}
 	G^R_0(\rr,t;\rr',t') &=-i  \langle \phi_s(\rr,t)\phi_a(\rr',t')\rangle = G^{sa}_0(\rr,t;\rr',t')\\
 	 G^K_0(\rr,t;\rr',t') &=-i  \langle \phi_s(\rr,t)\phi_s(\rr',t')\rangle = G^{ss}_0(\rr,t;\rr',t'),
 	\end{split}
 \end{equation}
 where the retarded Green's function $G^R_0$ and Keldysh Green's function $G^K_0$ describe the evolution of the system. Since $S_{ent}$ has a coupling between the $\phi_s$ and $\pi_s$ fields, it is useful to define $2 \times 2$ matrices of correlators
\begin{equation}
	\label{Greensfn}
	i\mathcal{G}^{pq}(\mathbf{r},t;\mathbf{r}',t') =
	{\small
		\begin{bmatrix}
			\langle \pi_p(\mathbf{r},t)\phi_q(\mathbf{r}',t')\rangle &
			\langle \pi_p(\mathbf{r},t)\pi_q(\mathbf{r}',t')\rangle \\
			-\langle \phi_p(\mathbf{r},t)\phi_q(\mathbf{r}',t')\rangle &
			-\langle \phi_p(\mathbf{r},t)\pi_q(\mathbf{r}',t')\rangle
	\end{bmatrix}}
\end{equation}
  where $p,q\in\lbrace s,a\rbrace$. Here $\mathcal{G}^{aa}$ is zero by the usual structure of SK field theory. These matrix propagators are represented by double lines as shown in Fig.~\ref{Free_theory_diagrams}(a). We denote the $2 \times 2$ Keldysh correlator matrix ($\mathcal{G}^{ss}$) by full solid double lines, whereas for retarded ($\mathcal{G}^{sa}$) and advanced ($\mathcal{G}^{as}$) correlator matrices we use  ``half-solid-half-dashed" and ``half-dashed-half-solid" double lines, respectively. Note that the components of these correlator matrices are not independent: $ \langle \phi_p(\rr,t)\pi_q(\rr',t') \rangle = \partial_{t'}\langle \phi_p(\rr,t)\phi_q(\rr',t') \rangle$, $ \langle \pi_p(\rr,t)\phi_q(\rr',t') \rangle = \partial_{t}\langle \phi_p(\rr,t)\phi_q(\rr',t') \rangle$ and $ \langle \pi_p(\rr,t)\pi_q(\rr',t') \rangle = \partial_t\partial_{t'}\langle \phi_p(\rr,t)\phi_q(\rr',t') \rangle$.  However, they provide a compact notation for calculating entanglement entropy. In thermal equilibrium (or any state with time reversal invariance), the off-diagonal elements of the $2 \times 2$ Keldysh correlator ($\mathcal{G}^{ss}$) vanish at equal time. 
 
 The quadratic term in $\mathcal S_{ent}$ can be treated as a replica off-diagonal self-energy represented by an ``Entanglement vertex" shown in Fig.~\ref{Free_theory_diagrams}(b). The vertex $\mathbb L$ couples symmetric fields $\phi_s$ and their conjugate momenta $\pi_s$ across replicas only inside the subsystem $A$ at time $t_0$. The $n^{\mathrm{th}}$ order R\'enyi entropy (for odd $n$) is the free energy of $n$ replicas in presence of $\mathcal S_{ent}$. This free energy can be written in terms of connected Feynman diagrams with the propagators $i\mathcal{G}^{ss}(\rr,t_0;\rr',t_0)$ and the entanglement vertices $(i\mathbb{L})$ where all lines are contracted into loops as shown in Fig.~\ref{Free_theory_diagrams}(c). Here, the spatial indices will run only over subsystem $A$, and the correlators are diagonal in replica space. Additionally, each diagram would require a trace over space in $A$, replica and $2\times 2$ space of the correlator matrices. 
 
 It is easy to show that the diagrams with an odd number of entanglement vertices vanish as $\mathbb{L}$ is skew-symmetric in replica space. The first non trivial diagram in Fig.~\ref{Free_theory_diagrams}(c) (second order in $\mathbb L$), evaluates to 
\begin{equation}
	\frac{1}{4}~\operatorname{Tr}\left[\mathbb{L}^2 \int_A d^d\rr\int_A d^d\rr' \mathcal G^{ss}(\rr,t_0,\rr',t_0)\mathcal G^{ss}(\rr',t_0,\rr,t_0)\right].
\end{equation}
Summing up all the diagrams in Fig.~\ref{Free_theory_diagrams}(c), we get 
\begin{equation}
	S_0^{(n)}=-V_A\ln 2+\frac{1}{2(n-1)} \operatorname{Tr}\ln\left[1+\mathbb{L}{\mathcal G}^{ss}\right].
	\label{S0n}
\end{equation}
Here, the trace is over replica, subsystem $A$, and $2\times 2$ space of correlator matrices. Using the fact that the eigenvalues of $\mathbb{L}$ are of the form $\lambda_k=i\tan(\pi k/n)$ where k runs from $1$ to odd integer $n$, and  evaluating  $\prod_{k=1}^n(1+\lambda_k\mathcal{G}^{ss})$, we can analytically compute the trace over the replica indices to obtain (for the details see Appendix \Ref{determinant_calculation})     
\begin{equation}\label{S_0neqn}
	\begin{split}
		S_0^{(n)}  &= \frac{1}{2(n-1)} \operatorname{Tr} \left( \ln \left[ \left(\frac{\mathcal{G}^{ss}+ 1}{2} \right)^n-\left(\frac{\mathcal{G}^{ss}- 1}{2} \right)^n \right]\right),
	\end{split}
\end{equation}
 where the trace is over the spatial indices in the subsystem $A$ as well as in the $2 \times 2$ space of the correlators. Note that since $n$ is odd, only even powers of $\mathcal{G}^{ss}$ survive in Eq.~\eqref{S_0neqn}~consistent with the fact that the diagrams with odd powers of $\mathbb L$  vanish due to the skew-symmetric nature of $\mathbb L$. 
 
 We can now analytically continue the odd R\'enyi entropies to $n\rightarrow 1$ to obtain the von Neumann entropy,
\begin{equation}\label{S_0VN_general_formula}
	S_0^{\mathrm{vN}}= \frac{1}{2}\,\operatorname{Tr}\!\left[ \frac{\mathcal{G}^{ss}+ 1}{2}\, \ln\frac{\mathcal{G}^{ss}+ 1}{2}-\frac{\mathcal{G}^{ss}- 1}{2} \, \ln\frac{\mathcal{G}^{ss}- 1}{2}\right].
\end{equation}

Finally, we consider the special case where the system is in thermal equilibrium. Here $\mathcal{G}^{ss}$ is purely off-diagonal.   Using  $M^2(\rr,t_0,\rr',t_0) = \int_{A}  d^d\rr_1 \langle \phi_s(\rr,t_0)\phi_s(\rr_1,t_0)\rangle~\langle \pi_s(\rr_1,t_0) \pi_s(\rr',t_0)\rangle $, we get (see Appendix~\Ref{Ground_state_case}) for details)
\begin{equation}\label{S_0neqn_equilibrium}
	\begin{split}
		S_0^{(n)}  &= \frac{1}{n-1} \operatorname{Tr}_A \left( \ln \left[ \left(\frac{M+ 1}{2} \right)^n-\left(\frac{M- 1}{2} \right)^n \right]\right),
	\end{split}
\end{equation}
 where the trace is now restricted to the spatial indices in $A$. We can now analytically continue the odd R\'enyi s to $n\rightarrow 1$ to obtain the von Neumann entropy,
\begin{equation}
	\begin{split}
		S_0^{\mathrm{vN}}&= \operatorname{Tr}_A\left[ \frac{\left(M+ 1\right)}{2} \ln\frac{\left(M+ 1\right)}{2}-\frac{\left(M- 1\right)}{2} \ln\frac{\left(M- 1\right)}{2}\right]. \\
	\end{split}
\end{equation}
These formulae match exactly with the results obtained previously \cite{Bombelli,Casini_2009} for the ground state of free scalar fields using completely different approaches. This provides an important check on our formalism. We have obtained a generalization of the well-known formulae for entanglement entropy of Gaussian scalar fields in terms of correlation functions, which applies to systems out of equilibrium. This allows us to consider dynamics~\cite{Mursalin_dipole} under quantum quenches~\cite{Alba_calabrese, Cotler2016,Sumit_Das_quench,banerjee2020quantum} as well as open quantum system dynamics of scalar fields coupled to baths~\cite{Mursalin2022}. We note that the infinite size of the local Hilbert space of the Bosons makes a numerical calculation of EE impossible to track (especially out of equilibrium), and we have now provided an exact formula to capture EE in this situation.

In the next section, we will consider a theory of interacting scalar fields and provide a diagrammatic construction of entanglement entropy for the interacting theory.

\section{\label{Interactions}Entanglement Entropy of Interacting Scalar fields}

We now extend our formulation to a theory of interacting scalar fields. We will consider a $\phi^4$ theory with local interactions as a prototype of an interacting theory (although the general principle works for more complicated theories, e.g., those with dipole conservation~\cite{Pretko_dipole,Mursalin_dipole}). The system is described by an action $\mathcal{S}=\mathcal{S}_{0}+\mathcal{S}_{\mathrm{int}}$, where $\mathcal{S}_0$ is given by Eq.~\eqref{freescalar} and
\beq
\mathcal{S}_{\mathrm{int}} =-\frac{2\lambda}{4!} \int d^d\rr \int dt~\left[\phi_s^3(\rr,t)\phi_a(\rr,t) + \phi_a^3(\rr,t)\phi_s(\rr,t)\right].
\eeq
This corresponds to a $ \lambda\phi^4/4!$ interaction on the forward and backward time contours, respectively. The R\'enyi entanglement entropy is then the free energy of the theory with the action $\sum_{\alpha=1}^n\mathcal{S}_0^{(\alpha)}+\mathcal{S}_{ent}+\sum_{\alpha=1}^n \mathcal{S}^{(\alpha)}_{\mathrm{int}}$. We will treat $\tilde{\mathcal{S}}=\sum_{\alpha=1}^n\mathcal{S}^{(\alpha)}_{0}+\mathcal{S}_{ent}$ as the Gaussian theory around which we expand our answers as an exact series of Feynman diagrams. Truncating the series upto a particular order in $\lambda$  will generate a perturbation theory for $S^{(n)}= S_0^{(n)}+ S_1^{(n)}+ ...$, where $S_l^{(n)} \sim \lambda^l$, and $S_0^{(n)}$ is given by Eq.~\eqref{S_0neqn}. We will leave the question of non-perturbative approximations for a later work. 

\begin{figure*}[t] \includegraphics[width=\textwidth]{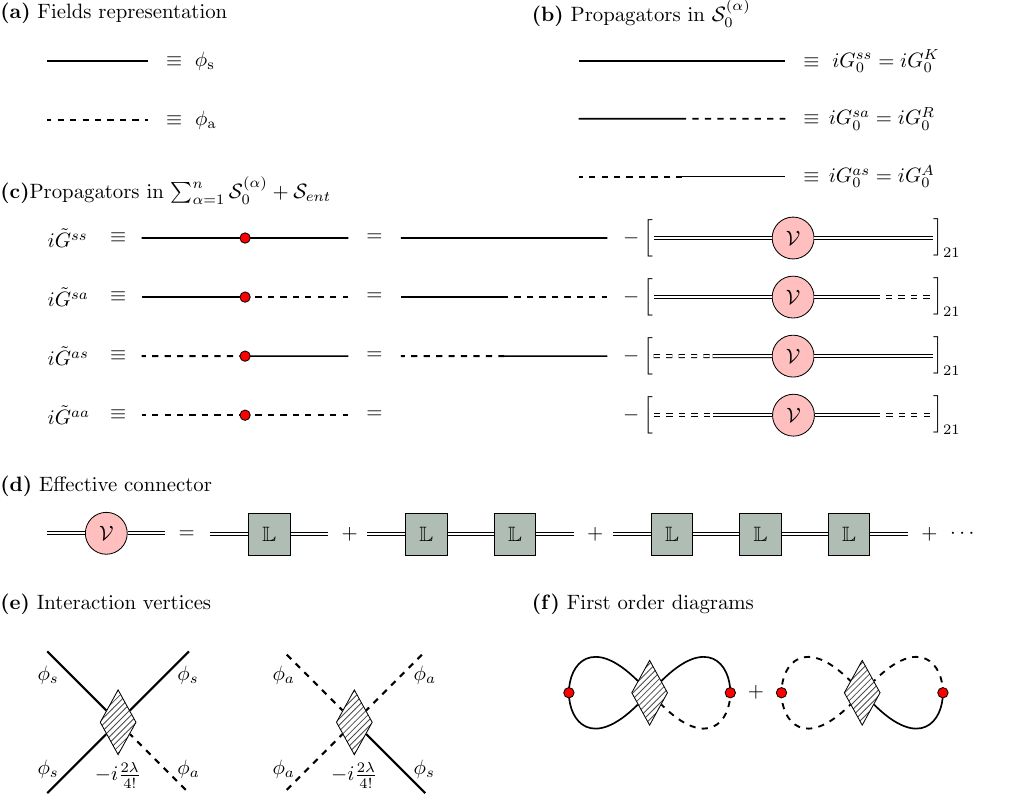}
	\caption{(a) The symmetric component $\phi_s$ is represented by a half solid line, and the antisymmetric component $\phi_a$ by a half dashed line. (b) The Keldysh $(G_0^{ss/K})$, retarded $(G_0^{sa/R})$  and advanced $(G_0^{as/A})$ propagators in the free theory $\mathcal{S}_{0}$ are denoted by a ``solid-solid", ``solid-dashed" and ``dashed-solid" line respectively and are replica-independent. (c) The propagators of the theory governed by $\tilde{\mathcal{S}}=\sum_{\alpha=1}^n\mathcal{S}^{(\alpha)}_{0}+\mathcal{S}_{ent}$ include contributions from both the free theory propagators and $\mathcal{S}_{ent}$. 
	The corresponding propagators $\tilde{G}^{ss}$, $\tilde{G}^{sa}$, $\tilde{G}^{as}$, and $\tilde{G}^{aa}$ of $\tilde{\mathcal{S}}$ are represented by ``solid–solid", ``solid–dashed", ``dashed–solid", and ``dashed–dashed" lines, respectively, each marked with a red circle at the centre indicating corrections from the effective connector $(\mathcal{V})$ defined in (d), obtained by resumming contributions from $\mathcal{S}_{ent}$. These propagators are replica-dependent. (e) The interaction term $\mathcal{S}_{\mathrm{int}}$ is represented by two four-leg vertices: one with three solid and one dashed line, and the other with three dashed and one solid line. Solid and dashed lines correspond to $\phi_s$ and $\phi_a$, respectively. Correction to $S^{(n)}$ due to interaction can be expressed as a sum of connected diagrams for various powers of interaction strength. The first-order corrections to $S^{(n)}$ are shown in (f).}
	\label{GtildepropagandFreeenergy}
\end{figure*}

The first task in this scheme is to find the propagators $\tilde{G}$ [represented by single lines with red circles at the centre, shown in Fig.~\ref{GtildepropagandFreeenergy}(c)] in the Gaussian theory described by $\tilde{\mathcal S}$. We use the $2\times2$ propagator matrices shown in Fig.~\ref{Free_theory_diagrams}(a) together with the entanglement vertex depicted in Fig.~\ref{Free_theory_diagrams}(b). Instead of the free energy diagrams of Fig.~\ref{Free_theory_diagrams}(c), we resum the Dyson series for the propagator as shown in Fig.~\ref{GtildepropagandFreeenergy}(d). This defines an effective connector $\mathcal{V}$ that encodes the effect of $\mathcal S_{ent}$. Note that the series in Fig.~\ref{GtildepropagandFreeenergy}(d) involve only the $2\times 2$ Keldysh propagators $(\mathcal{G}^{ss})$, i.e, $\mathcal{V}$ connects symmetric fields and their conjugate momenta between replicas. Finally, the modified correlators $\tilde{G}$ are given by the diagrams in Fig.~\ref{GtildepropagandFreeenergy}(c) in terms of the original free propagators $G_0^{K}$, $G_0^{R}$, and $G_0^{A}$, represented by ``solid-solid", ``solid-dashed", and ``dashed-solid" lines, respectively, in Fig.~\ref{GtildepropagandFreeenergy}(b), and the connector $\mathcal{V}$ shown in Fig.~\ref{GtildepropagandFreeenergy}(d). The connector $\mathcal{V}$  plays a role analogous to the $T$ matrices in scattering problems. In fact, the entanglement cut acts as a source of inter-replica scattering. Thus, $\tilde{G}$ can be thought of as propagators in the presence of inter-replica scatterings induced inside the entanglement cut.

The propagators $\tilde{G}$ will, in general, be off-diagonal in replica indices. Due to the entanglement cut, translation invariance is no longer a good symmetry even in a system, while the original action is translation invariant. Since the interactions couple only $\phi$ fields, we will only need the $\langle \phi\phi\rangle$ correlators in $\tilde{\mathcal{S}}$ to construct the perturbation series. The presence of $\mathcal{S}_{ent}$ implies that $\tilde{G}^{aa}$ correlators, which vanish by construction in standard Keldysh field theories, will be non-zero in this case. In addition, $\tilde{G}^{sa}$ will not have the standard retarded structure. As a result, we will indicate the $\tilde{G}$ Green's functions by putting Keldysh indices and refrain from using terms like retarded correlators in this section.  One can show that  $\tilde{G}^{ss}$ and $\tilde{G}^{aa}$ are still anti-Hermitian in their indices. As shown in Fig.~\ref{GtildepropagandFreeenergy}(c), the correlators are given by
\begin{widetext}
	\bqa
	\begin{split}
		\bar{G}^{sa}_{\alpha\beta}(\rr,t;\rr',t')&=G_0^{sa}(\rr,t;\rr',t')\delta_{\alpha\beta}- i\Big[\int_A d^d\rr_1\int_Ad^d\rr_2~ \mathcal{G}^{ss}(\rr,t;\rr_1,t_0) \mathcal{ V}_{\alpha\beta}(\rr_1,\rr_2)\mathcal{G}^{sa}(\rr_2,t_0;\rr't')\Big]_{21}\\
		\tilde{G}^{as}_{\alpha\beta}(\rr,t;\rr',t')&=G_0^{as}(\rr,t;\rr',t')\delta_{\alpha\beta}- i\Big[\int_A d^d\rr_1\int_Ad^d\rr_2 \mathcal {G}^{as}(\rr,t;\rr_1,t_0) \mathcal{ V}_{\alpha\beta}(\rr_1,\rr_2)\mathcal {G}^{ss}(\rr_2,t_0;\rr't')\Big]_{21}\\
		\tilde{G}^{ss}_{\alpha\beta}(\rr,t;\rr',t')&=G_0^{ss}(\rr,t;\rr',t')\delta_{\alpha\beta}- i\Big[\int_A d^d\rr_1\int_Ad^d\rr_2 \mathcal {G}^{ss}(\rr,t;\rr_1,t_0) \mathcal{ V}_{\alpha\beta}(\rr_1,\rr_2)\mathcal {G}^{ss}(\rr_2,t_0;\rr't')\Big]_{21}\\
		\tilde{G}_{\alpha\beta}^{aa}(\rr,t;\rr',t')&= -i\Big[\int_A d^d\rr_1\int_Ad^d\rr_2 \mathcal {G}^{as}(\rr,t;\rr_1,t_0) \mathcal{ V}_{\alpha\beta}(\rr_1,\rr_2)\mathcal {G}^{sa}(\rr_2,t_0;\rr't')\Big]_{21};
	\end{split}
	\eqa
\end{widetext}
where 
\begin{equation}
		\hat{\mathcal{V}} =  i\Big[1+\mathbb{L}\mathcal{G}^{ss}\Big]^{-1}\mathbb{L},
\end{equation}
and $G_0$ are the corresponding propagators in the non-interacting theory in a single replica (they are diagonal in replica indices). Here, $\left[\dots\right]_{21}$ denotes the $(2,1)$ element of a $2\times2$ matrix.  Note that while $G_0^{aa}$ is $0$, $\tilde{G}^{aa}$ is not.  $\hat{\mathcal{V}}$ has a circulant structure in the replica space of the form (see Appendix \Ref{V_matrix_derivation} for details)-
\begin{equation}
	\hat{\mathcal{V}} =\left(\begin{array}{ccccccccc}
		v_0 & v_{n-1}& v_{n-2} & \cdot	&\cdot &\cdot &\cdot &v_{2} &v_{1}\\
		v_1 & v_0& v_{n-1} & v_{n-2}	&\cdot &\cdot &\cdot &\cdot &v_{2}\\
		v_2 & v_1& v_0 & v_{n-1}	&\cdot &\cdot &\cdot &\cdot &v_{3}\\
		\cdot & v_2& v_1 &v_0&\cdot &\cdot &\cdot &\cdot &\cdot\\
		\cdot &\cdot &\cdot &\cdot&\cdot &\cdot &\cdot &\cdot&\cdot\\
		\cdot &\cdot &\cdot &\cdot&\cdot &\cdot &\cdot &\cdot&\cdot\\
		\cdot &\cdot &\cdot &\cdot&\cdot &\cdot &\cdot &\cdot&\cdot\\
		v_{n-2} &\cdot &\cdot &\cdot&\cdot &\cdot &\cdot &v_0&v_{n-1}\\
		v_{n-1} &v_{n-2} &v_{n-3}&\cdot&\cdot &\cdot &\cdot &v_{1}&v_0\\
	\end{array}\right)_{n\times n};
\end{equation}	
where $v_k$'s are matrices given by 
\begin{eqnarray}
	\nonumber	v_0&=&i\left[\frac{(\mathcal{G}^{ss}+1)^{n-1}-(\mathcal{G}^{ss}-1)^{n-1}}{(\mathcal{G}^{ss}+1)^{n}-(\mathcal{G}^{ss}-1)^{n}}\right]\\
	v_{k\geq 1}&=&-2i\left[\frac{ (\mathcal{G}^{ss}+1)^{n-k-1}(\mathcal{G}^{ss}-1)^{k-1}}{ (\mathcal{G}^{ss}+1)^{n}-(\mathcal{G}^{ss}-1)^{n}}\right].
\end{eqnarray}

Since entanglement entropy is the free energy of the replicated theory with inter-replica couplings, it is now straightforward to construct a perturbative expansion of $S^{(n)}$ of the interacting theory. The microscopic interaction terms are represented by the vertices shown in Fig.~\ref{GtildepropagandFreeenergy}(e). Here, the solid lines indicate symmetric fields and dashed lines indicate antisymmetric fields. There are two different vertices: one with three symmetric ($\phi_s$) and one antisymmetric ($\phi_a$) field, and the other with three antisymmetric ($\phi_a$) and one symmetric ($\phi_s$) field, as in standard Keldysh field theory for scalar Bosons. Note that the interaction terms are replica diagonal; i.e., all fields emanating from a vertex must carry the same replica index. This will constrain the replica structure of the terms in perturbation theory. The perturbation series is best represented by free energy Feynman diagrams  (i.e., connected diagrams where all lines are connected in loops, leaving no hanging endpoints). Fig.~\ref{GtildepropagandFreeenergy}(f) shows the diagrams for $1^{\mathrm{st}}$ order corrections, while in Appendix~\ref{Second_order_diagrams}, Fig.~\ref{Second_order_Free_energy_Diagrams}(i)-(xix) shows all possible $2^{\mathrm{nd}}$ order diagrams. While the diagrams look like standard free energy diagrams and the rules for evaluating them are similar, a few things need to be kept in mind: 
\begin {itemize}

\item All single lines with a red circle in the centre represent $\tilde{G}$ propagators which carry replica indices in addition to space-time indices. The propagator $\tilde{G}$  can be off-diagonal in replica indices.

\item Unlike standard Keldysh field theory, $\tilde{G}_{aa} \neq 0$, and hence there will be additional free energy diagrams at each order.

\item All lines coming from an interaction vertex must carry the same replica index, and this constraint is implicit in the diagrams.

\item Evaluation of the diagrams involves: \begin{enumerate}  
	\item A factor of $i\tilde{G}$ for each red circled propagator.
	
	\item A factor of $-i\frac{2\lambda}{4!}$ for each vertex.
	
	\item A symmetry factor to take care of different contractions that yield the same diagram.
	
	\item Integration over all space-time indices {\it(space is not restricted to subsystem A)} and trace over all replica indices.
	\end{enumerate}
\end{itemize}

As an example, the ${\cal O}(\lambda)$  corrections to $S^{(n)}$ shown in Fig.~\ref{GtildepropagandFreeenergy}(f) evaluates to
\begin{equation}
	\begin{split}
		=-\frac{i\lambda}{4}\sum_{\alpha=1}^{n}&\int d^d\rr \int dt~\left[i\tilde{G}^{ss}_{\alpha\alpha}(\rr,t;\rr,t)+i\tilde{G}^{aa}_{\alpha\alpha}(\rr,t;\rr,t)\right]\\
		&\times\frac{1}{2}\left[i\tilde{G}^{sa}_{\alpha\alpha}(\rr,t;\rr,t)+i\tilde{G}^{as}_{\alpha\alpha}(\rr,t;\rr,t)\right].
	\end{split}
\end{equation}
The corrections for ${\cal O}(\lambda^2)$ are shown in Appendix~\Ref{Second_order_diagrams} in detail. The continuum theory should be supplemented with a UV regulation (such as a cutoff $\Lambda$) to make these integrals finite. Note that entanglement entropy is known to depend on UV cutoffs in an essential way (at least for free field theories).

Finally, we would like to comment on some simplifications that occur when we take the limit $n\rightarrow 1$ to calculate $S^{\mathrm{vN}}$. The main point to note is $v_0 = \mathcal{V}_{\alpha\alpha} \sim (n-1)$. Since we divide the free energy by $n-1$ before taking the limit of $n \rightarrow 1$, the contribution from a diagram with more than one factor of $v_0$ will vanish. For example, in the first order diagram shown in Fig.~\ref{GtildepropagandFreeenergy}(f), all the propagators have indices $(\alpha,\alpha)$.  Note that, since $\tilde{G}^{aa} \sim (n-1)$; and $G_0^{sa}(t,t)+G_0^{as}(t,t)=0$, one can easily show that the term involving $\tilde{G}^{aa}$ will have a vanishing contribution to the von Neumann entanglement entropy. Hence, the first order contribution to $S^{\mathrm{vN}}$ is then given by
\begin{widetext}
	\begin{equation}
		\begin{split}
			\delta S^{\mathrm{vN}}&=\frac{i\lambda}{8}\int_A d^d\rr \int dt~iG^K_0(\rr,t;\rr,t)\\
			&\times\int_A d^d\rr_1\int_A d^d\rr_2
			\Big[i\mathcal{G}^{ss}(\rr,t;\rr_1,t_0)\tilde{v}(\rr_1,\rr_2)i\mathcal{G}^{sa}(\rr_2,t_0;\rr,t)+i\mathcal{G}^{as}(\rr,t;\rr_1,t_0)\tilde{v}(\rr_1,\rr_2)i\mathcal{G}^{ss}(\rr_2,t_0;\rr,t)\Big]_{21},
		\end{split}
	\end{equation}
\end{widetext}
where 
\begin{equation}
	\tilde{v} =\lim_{n\to1}\frac{v_0}{1-n}=-\frac{i}{2} \Big(\ln \left[ \mathcal{G}^{ss}+1\right]-\ln\left[\mathcal{G}^{ss}-1\right]\Big).
\end{equation}
We note that, in contrast to a theory of interacting fermions, where the first-order correction to $S^{\mathrm{vN}}$ vanishes, the corresponding correction in an interacting scalar field theory is, in general, nonzero. Here, the first-order terms generate a correction to the mass of the fields, which changes the quantization and hence the distribution of the particles. This leads to a correction to the entanglement entropy of the system. 
	
\section{\label{Conclusions}Conclusions}
	
	In this paper, we have constructed a quantum field theory for calculating entanglement entropy of interacting real scalar fields both in and out of equilibrium. We show that the odd order R\'enyi entropies can be written as the free energy of replicas with a quadratic inter-replica coupling.  This inter-replica coupling replaces the field matching conditions in a standard field theory and leads to the construction of a detailed dictionary between correlators (in a single replica) and entanglement entropy. For a free theory, we obtain exact formulae for an open quantum system undergoing non-equilibrium dynamics. For an interacting theory, we provide a dictionary of Feynman diagrams and work out the first 2 orders in a perturbation theory in coupling. This extends the formalism developed by Chakraborty and Sensarma \cite{AhanaPRL,AhanaPRA} for Schr\"odinger Bosons and Moitra and Sensarma for Fermions~\cite{moitra2020entanglement,Saranyo_Building_Entanglement}. 
	
	We note that while we have shown how to construct all possible Feynman diagrams for calculating the entanglement entropy of the interacting system, a practical calculation for a given situation would involve making approximations appropriate to the corresponding problem. We have explicitly considered a perturbation theory in the coupling constant up to $2^{\mathrm{nd}}$ order. One can also consider non-perturbative approximations, such as renormalized perturbation theory or large $N$ expansions, which we have not considered in this paper. All of these are equivalent to selecting a subset of the Feynman diagrams and resumming them to all orders. A detailed investigation of these non-perturbative approximations and their applications is left for future work.
	
	\medskip
	
	
	\begin{acknowledgments}
		
		M.K.S. and R.S. acknowledge funding from the Department of Atomic Energy, Govt. of India. The authors acknowledge useful discussions with Prof. Onkar Parrikar, Prof. Subir Sachdev, Prof. Gautam Mandal, and Prof. Sandip Trivedi.
	\end{acknowledgments}
	
	\onecolumngrid
	\appendix

\newpage 

\section{Details of Gaussian integrals over $\mathcal{J}$ to obtain $\mathcal{S}_{ent}$}\label{J_integrals_for_odd_and_even_Renyi}

In the main text, we have obtained a formula for the $n^{\mathrm{th}}$ order R\'enyi entropy,
\begin{equation}
	\begin{split}
		e^{(1-n)S^{(n)}}&= \int \prod_{\alpha=1}^n {\cal D}[\phi^{(\alpha)}_{s,a}]~ \exp\left[i\sum_{\alpha=1}^n \mathcal{S}[\phi^{(\alpha)}_{s,a}]\right]\\
		&\times\int \prod_{\alpha=1}^{n-1} ~{\mathscr D}[{\bf J^{(\alpha)}}]~\exp\left[i\sum_{\alpha=1}^{n-1} \int_A d^d\rr ~J_\rr^{\phi(\alpha)} \left[\phi^{(\alpha)}_s(\rr,t_0)-\phi^{(n)}_s(\rr,t_0)\right]-J_\rr^{\pi(\alpha)} \left[\pi^{(\alpha)}_s(\rr,t_0)-\pi^{(n)}_s(\rr,t_0)\right]\right]\\
		&\times~ \exp\left[\frac{i}{2}{\cal J}^T\mathbb{K}_{n-1}\otimes \left[\begin{array}{cc} 0 &-1\\
			1 & 0 \end{array}\right]\otimes\hat{P}_A{\cal J}\right],\\
	\end{split}
\end{equation}
where $\phi^{(\alpha)}$ are the replica fields and ${\cal J}^T=\left(\JJ^{\phi(1)},\JJ^{\pi(1)},..\JJ^{\phi(n-1)},\JJ^{\pi(n -1)}\right)$ are the sources.	
Here we will perform Gaussian integrals over the sources $\mathcal{J}$, which would involve the inverse of  $\mathbb K_{n-1}\otimes(-i\sigma_y)\otimes\hat{P}_A$, given by $\mathbb K^{-1}_{n-1}\otimes(i\sigma_y)\otimes\hat{P}_A$. We will now focus on the eigenvalues and eigenvectors of the matrix 
\beq
\mathbb{K}_{n-1}=\left( \begin{array}{ccccc}
	0 & -1 & -1& ... & ...\\
	1 &0 &  -1 & -1& ...\\
	1 &1 & 0 & -1& ...\\
	... & 1 & 1& 0& ...\\
	... & ... & ...& ...& ...
\end{array}\right).
\eeq 
Since $\mathbb{K}_{n-1} $ is a skew circulant matrix, its eigenvectors must be of the form 
\beq\label{evec_K}
|k\rangle = \frac{1}{\sqrt{n-1}}(\omega_k,\omega_k^2,...\omega_k^{n-1})^T,
\eeq
with $\omega_k^{n-1}=-1$, i.e., $\omega_k=e^{i\frac{\pi (2k-1)}{n-1}}$, where $k=1,2,...,n-1$. Putting this into the eigenvalue equation, it is easy to show that the eigenvalues are given by 
\beq\label{eval_K}
\epsilon_k = \frac{\omega_k+1}{\omega_k-1}=-i\cot\left(\frac{\pi (2k-1)}{2(n-1)}\right).
\eeq

\subsection{Calculation of odd order R\'enyi Entropy $S^{(n)}$, where $n=(2q+1)$}\label{A_odd_Renyi}

In this case $\mathbb K_{n-1}$ is invertible and $\mathbb{L}_{n-1}= \mathbb{K}^{-1}_{n-1}$ is given by 
\begin{equation}
	\begin{split}
		\label{eqLmatrix}
		\mathbb{L}^{\alpha\beta}_{n-1}=\frac{1}{n-1}\sum_{k=1}^{n-1} \frac{\omega_k^{\alpha-\beta}}{\epsilon_k}=\frac{1}{n-1}\sum_{k=1}^{n-1} \frac{\omega_k^{\alpha-\beta} \left(\omega_k-1\right)}{\left(\omega_k+1\right)}=\frac{-1}{n-1}\sum_{k=1}^{n-1} \frac{\omega_k^{n-1+\alpha-\beta} \left(\omega_k-1\right)}{\left(\omega_k+1\right)}
	\end{split}
\end{equation}
where we have used $\omega_k^{n-1}=-1$ to get the last expression. We evaluate the sum in Eq.~\eqref{eqLmatrix} by relating it to a contour integral. The meromorphic complex function
\begin{equation}
	f(z)=\frac{z^{p-1}}{(z^{n-1}+1)(z+1)}~~\text{~~for~~} 0\leq p\leq n-1; 
\end{equation}
has simple poles at $z=\omega_k$ and $z=-1$. Note that for $p=0$, $f(z)$ will have an additional simple pole at $z=0$. For a circular contour $(C)$ of large radius $R\rightarrow \infty$, traversed anticlockwise,
\begin{equation}\label{contint}
	\oint_{C}f(z)~dz= 2\pi i\left[-\sum_{k=1}^{n-1}\frac{\omega_k^p}{(n-1)(\omega_k+1)}+ \delta_{p0} +\frac{(-1)^{p-1}}{(-1)^{n-1}+1}\right] =0.
\end{equation}
Here, the first expression uses Cauchy's residue theorem, while the last expression uses the fact that for $|z| \rightarrow \infty$, $f(z) \sim 1/z^{2+n-1-p}$, and the contour integral vanishes for $0 \leq p \leq n-1$. Hence, we have
\begin{equation}
	\frac{1}{n-1}\sum_{k=1}^{n-1}\frac{w_k^{p}}{w_k+1}=\delta_{p0}+\frac{(-1)^{p-1}}{1+(-1)^{n-1}}; ~~\forall ~~0\leq p\leq n-1.
\end{equation} 
Using the above formula to evaluate the sums in Eq.~\eqref{eqLmatrix} for odd $n$, we get
\begin{eqnarray}
	\mathbb{L}^{\alpha\alpha}_{n-1}&=&\frac{1-(-1)^{n-1}}{1+(-1)^{n-1}}=0 \\
\nonumber	\mathbb{L}^{\alpha\beta}_{n-1}&=&\frac{2(-1)^{\alpha-\beta}}{1+(-1)^{n-1}}~=~(-1)^{\alpha-\beta}~~~ \forall ~~\alpha >\beta\\
\nonumber	\mathbb{L}^{\alpha\beta}_{n-1}&=&~-~(-1)^{\alpha-\beta}~~~ \forall ~~\beta >\alpha
\end{eqnarray}
i.e.,  $\mathbb{L}_{n-1}^{\alpha\beta}=[\mathbb{K}_{n-1}^{-1}]^{\alpha\beta}=(-1)^{\alpha-\beta}[\Theta(\alpha-\beta)-\Theta{(\beta-\alpha)}]$. 

Using the Inverse of $\mathbb K_{n-1}$, the Gaussian integral over $\mathcal J$ can be done to obtain the extra term
\begin{equation}
	\label{App:Sent:odd}
	\begin{split}
	\mathcal{S}_{ent} &=\frac{1}{2}\sum_{\alpha,\beta=1}^{n-1}\int_A d^d \rr~ \left[\begin{array}{c}\phi^{(\alpha)}_s(\rr,t_0)-\phi^{(n)}_s(\rr,t_0),\pi^{(\alpha)}_s(\rr,t_0)-\pi^{(n)}_s(\rr,t_0)\\ \end{array}\right]\mathbb{L}^{\alpha\beta}_{n-1}\left[\begin{array}{c}   \pi^{(\beta)}_s(\rr,t_0)-\pi^{(n)}_s(\rr,t_0)\\-(\phi^{(\beta)}_s(\rr,t_0)-\phi^{(n)}_s(\rr,t_0)) \end{array}\right]\\
		&=\frac{1}{2}\sum_{\alpha,\beta=1}^{n}\int_A d^d \rr~ \left[\begin{array}{c}\phi^{(\alpha)}_s(\rr,t_0),\pi^{(\alpha)}_s(\rr,t_0)\\ \end{array}\right]\mathbb{L}^{\alpha\beta}_{n}\left[\begin{array}{c}   \pi^{(\beta)}_s(\rr,t_0)\\-\phi^{(\beta)}_s(\rr,t_0) \end{array}\right],
	\end{split}
\end{equation} 
where 
\begin{equation}
	\mathbb{L}_n=
	\begin{bmatrix}
		\ \: \: \: 0 & \:\:\: 1 & -1 &\:\:\:1 &... &-1\\
		\ -1&  \: \: \:0 & \:\:\:1 &-1&... & \: \: \: 1\\
		\ \:\:\:1& -1  & \: \: \:0 &\:\:\:1&... &-1\\
		-1& \:\:\:1  & -1 &\: \: \:0&... &\: \: \: 1\\
		\ \: \: \: . & \: \: \: . & \: \: \: .&\: \: \: .& &  \: \: \: . \\
		
		\ \: \: \:1& -1 &\: \: \:1 &-1& ...& \: \: \:0 \\
	\end{bmatrix}_{n\times n}.
	\label{Ln}
\end{equation}
Here we have used $\mathbb{L}_n^{\alpha,\beta}=\mathbb{L}_{n-1}^{\alpha,\beta}$ for $1\leq\alpha,\beta\leq(n-1)$, $\mathbb{L}_n^{\alpha n}=-\sum_{\beta=1}^{n-1}\mathbb{L}^{\alpha\beta}_{n-1}=(-1)^\alpha$, $\mathbb{L}_n^{n\beta}=-\sum_{\alpha=1}^{n-1}\mathbb{L}^{\alpha\beta}_{n-1}=-(-1)^\beta$ and $\mathbb{L}_n^{n n}=\sum_{\alpha,\beta=1}^{n-1}\mathbb{L}^{\alpha\beta}_{n-1}=0$.
Eq.~\eqref{App:Sent:odd} is the same as Eq.~\eqref{eq:Sent:odd} shown in the main text, where the matrix $\mathbb{L}_n$ is written as $\mathbb{L}$.

\subsection{Calculation of even order R\'enyi Entropy $S^{(n)}$, where $n=2q$}\label{A_even_Renyi}

The eigenvalues and eigenvectors of $\mathbb{K}_{n-1}$ are still given by Eq.~\eqref{eval_K} and~\eqref{evec_K}; the big difference is that $\omega_{k=n/2}=-1$ and hence $\epsilon_{k=n/2}=0$, with the zero mode eigenvector given by $|n/2\rangle = (1,-1,1,-1,...1,-1)/\sqrt{n-1}$. This leads to 2 big changes vis a vis the calculation for odd order R\'enyi entropies: (i) integration over the zero mode leads to the boundary conditions: $\sum_\alpha (-1)^\alpha  \phi^{(\alpha)}_s(\rr,t_0) =0$ and  $\sum_\alpha (-1)^\alpha  \pi^{(\alpha)}_s(\rr,t_0) =0$ for all $\rr \in A$. (ii) Integrations over the remaining modes lead to the matrix $\mathbb{P}_{n-1}$, which is the Moore–Penrose pseudoinverse of $\mathbb{K}_{n-1}$. Thus, the inverse is replaced by the pseudoinverse. The pseudoinverse can be constructed from the non-zero eigenvalues and eigenvectors
\begin{equation}
	\mathbb{P}^{\alpha\beta}_{n-1}=\frac{1}{n-1}\sum_{k=1; k\neq\frac{n}{2}}^{n-1}\frac{\omega_k^{\alpha-\beta}}{\epsilon_k}~=~\frac{1}{n-1}\sum_{k=1; k\neq\frac{n}{2}}^{n-1}\frac{\omega_k^{\alpha-\beta}(\omega_k-1)}{(\omega_k+1)}.
\end{equation}
To evaluate these sums, we once again proceed by evaluating the contour integral in Eq.~\eqref{contint} in two ways: using the residue theorem and a direct evaluation. The key difference is that since $\omega_{n/2}=-1$, $f(z)$ has a double pole at $z=-1$ and the residues must be calculated keeping this in mind. All other singularities remain simple poles. This leads to
\begin{equation}
	\frac{1}{n-1}\sum_{k=1;k\neq\frac{n}{2}}^{n-1}\frac{\omega_k^{p}}{\omega_k+1}=\delta_{p0}+\frac{(-1)^{p-1}}{2(n-1)}(n-2p) ~~\text{~~for~~} 0\leq p\leq n-1.
\end{equation}
Using this, one obtains the following:
\begin{eqnarray}
	\mathbb{P}^{\alpha\alpha}_{n-1}&=&0\\
	\nonumber	\mathbb{P}^{\alpha\beta}_{n-1}&=&(-1)^{\alpha-\beta}\left[1-\frac{2(\alpha-\beta)}{n-1}\right]~~~ \forall ~~\alpha >\beta\\
	\nonumber	\mathbb{P}^{\alpha\beta}_{n-1}&=&~-~(-1)^{\alpha-\beta}\left[1-\frac{2(\beta-\alpha)}{n-1}\right]~~~~ \forall ~~\beta >\alpha
\end{eqnarray}i.e.,  $\mathbb{P}_{n-1}^{\alpha\beta}=(-1)^{\alpha-\beta}\left[\operatorname{sgn}(\alpha-\beta) - \dfrac{2(\alpha-\beta)}{\,n-1}\right]$.
Thus, we arrive at the entangling action
\begin{equation}
	\label{App:Sent:even}
	\begin{split}
		\mathcal{S}_{ent} &=\frac{1}{2}\sum_{\alpha,\beta=1}^{n-1}\int_A d^d \rr~ \left[\begin{array}{c}\phi^{(\alpha)}_s(\rr,t_0)-\phi^{(n)}_s(\rr,t_0),\pi^{(\alpha)}_s(\rr,t_0)-\pi^{(n)}_s(\rr,t_0)\\ \end{array}\right]\mathbb{P}^{\alpha\beta}_{n-1}\left[\begin{array}{c}   \pi^{(\beta)}_s(\rr,t_0)-\pi^{(n)}_s(\rr,t_0)\\-(\phi^{(\beta)}_s(\rr,t_0)-\phi^{(n)}_s(\rr,t_0)) \end{array}\right]\\
		&=\frac{1}{2}\sum_{\alpha,\beta=1}^{n}\int_A d^d \rr~ \left[\begin{array}{c}\phi^{(\alpha)}_s(\rr,t_0),\pi^{(\alpha)}_s(\rr,t_0)\\ \end{array}\right]\mathbb{P}^{\alpha\beta}_{n}\left[\begin{array}{c}   \pi^{(\beta)}_s(\rr,t_0)\\-\phi^{(\beta)}_s(\rr,t_0) \end{array}\right],
	\end{split}
\end{equation} 
where 
\begin{equation}
	\mathbb{P}^{\alpha\beta}_n =
	\begin{cases}
		(-1)^{\alpha-\beta}\left[\operatorname{sgn}(\alpha-\beta) - \dfrac{2(\alpha-\beta)}{\,n-1}\right], & 1 \leq \alpha,\beta \leq n-1, \\[1em]
		\frac{(-1)^{-\beta}}{n-1}\left(2\beta - n\right), & \alpha = n,\; 1 \leq \beta \leq n-1, \\[1em]
		-\frac{(-1)^{\alpha}}{n-1}\left(2\alpha - n\right), & 1 \leq \alpha \leq n-1,\; \beta = n.\\[1em]
		0, & \alpha = \beta = n.		
	\end{cases}
\end{equation}
This, together with the two boundary conditions, leads to Eq.~\eqref{Sneven} in the main text.

\section{Analytical derivation of $S_{0}^{(n)}$ for odd $n$ in a  Gaussian Theory }\label{determinant_calculation}

For odd $n$, $S_0^{(n)}$ is given by Eq.~\eqref{S0n} in the main text. Our goal is to find the analytical form of $\textrm{Tr}_\mathcal{R}\ln\left[1+\mathbb{L}\mathcal{G}^{ss}\right]$, where $\mathbb{L}$ is given by Eq.~\eqref{Ln} and $\mathcal{G}^{ss}$ is diagonal in replica space.  The eigenvectors of skew symmetric circulant matrix $\mathbb{L}_n$ are given by $|k\rangle =\frac{1}{\sqrt{n}}(r_k,r_k^2,...r_k^n)$ with $r_k^n=1$, i.e. $r_k = e^{i\frac{2\pi k}{n}}$ with $k=1,2,...n$. The corresponding eigenvalues  are 
\beq
\lambda_k =\frac{r_k-1}{r_k+1} =~i~\tan \frac{\pi k}{n}.
\eeq
To evaluate the replica trace, it is easier to replace the matrix $\mathcal{G}^{ss}$ (in real space and $\phi-\pi$ space) by one of its eigenvalues and then sum over these eigenvalues at the end. Assuming $g$ to be the eigenvalue of $\mathcal{G}^{ss}$, we obtain
\begin{equation}
	\begin{split}
		\operatorname{Tr}_\mathcal{R}\ln\left[1+\mathbb{L}g\right]&=\ln\left[\prod_{k=1}^{n}\left(1+ig\tan\frac{\pi k}{n}\right)\right]\\
		&=\ln\left[(g+1)^n\prod_{k=1}^{n}\left(\frac{e^{i\frac{2\pi k}{n}}-\frac{g-1}{g+1}}{e^{i\frac{2\pi k}{n}}+1}\right)\right].
	\end{split}
\end{equation}
Here, $\operatorname{Tr}_\mathcal{R}$ represents trace over replica indices. We know that for any complex variable $z$, the function $1-z^n$ can be written in terms of $n^{\mathrm{th}}$ roots of unity $r_k=e^{i\frac{2\pi k}{n}}$ as follows,
\begin{equation}
	\begin{split}
		1-z^n=\prod_{k=1}^n(r_k-z);
	\end{split}
\end{equation}
putting $z=(g-1)/(g+1)$ for the numerator and $z=-1$ for the denominator we have,
\begin{equation}
\operatorname{Tr}_\mathcal{R}\ln\left[1+\mathbb{L}g\right]=\ln\;(g+1)^n\left[\frac{1-\left(\frac{g-1}{g+1}\right)^n}{1-(-1)^n}\right]=\ln\left[\frac{(g+1)^n-(g-1)^n}{2}\right] ~~\text{~~for odd~~} n.
\end{equation}
Now, summing over all the eigenvalues $g$, we get
\begin{equation}
\operatorname{Tr}_\mathcal{R}\ln\left[1+\mathbb{L}\mathcal{G}^{ss}\right]=	\ln\left[\frac{(\mathcal{G}^{ss}+1)^n-(\mathcal{G}^{ss}-1)^n}{2}\right].
\end{equation}
This formula has been used in Eq.~\eqref{S_0neqn} in the main text.

\section{Derivation of $\mathcal{V}$ matrix for interacing Bosons}\label{V_matrix_derivation}

Here we will derive the matrix elements of the connector matrix $\hat{\mathcal{V}}=i(1+\mathbb{L}\mathcal{G}^{ss})^{-1}\mathbb{L}$ with the help of the eigenvalues and eigenvectors of $\mathbb{L}$ derived in Appendix \ref{determinant_calculation}. We will once again work with an eigenvalue $g$ of $\mathcal{G}^{ss}$, and sum over these eigenvalues at the end. We can do this because $\mathcal{G}^{ss}$ is diagonal in the replica space. We then have
\begin{equation}
	\begin{split}
	V_{\alpha\beta}(g)~=~i~\sum_{k=1}^{n}~(1+\lambda_kg)^{-1}\lambda_k r_k^{\alpha-\beta}~=~i~\frac{1}{n}\sum_{k=1}^{n}~\frac{\left(\frac{r_k-1}{r_k+1}\right)r_k^{\alpha-\beta}}{1+\left(\frac{r_k-1}{r_k+1}\right)g}~=~\frac{i}{(g+1)n}~\sum_{k=1}^{n}~\frac{r_k^{\alpha-\beta}(r_k-1)}{r_k-\left(\frac{g-1}{g+1}\right)}.
	\end{split}
\end{equation}
This sum can be evaluated using a contour integral trick similar to the one used in Appendix~\ref{J_integrals_for_odd_and_even_Renyi}, where we need to consider contour integrals of the function
\begin{equation}
	f(z)=\frac{z^{p-1}}{(z^n-1)(z-a)}. 
\end{equation}
This yields 
\begin{equation}
	\frac{1}{n}\sum_{k=1}^{n}\frac{r_k^{p}}{r_k-a}=-\left(\frac{a^{p-1}}{a^n-1}+\frac{\delta_{p0}}{a}\right); ~~~\forall ~0\leq p\leq n \text{~and~} a \notin\lbrace r_k\rbrace.
\end{equation} 
Using the above formula with $a=\frac{g-1}{g+1}$, the diagonal elements $\mathcal{V}_{\alpha\alpha}$ become
\begin{equation}
	\begin{split}
		V_{\alpha\alpha}(g)~&=~\frac{i}{(g+1)n}~\sum_{k=1}^{n}~\left[\frac{r_k}{r_k-\left(\frac{g-1}{g+1}\right)}-\frac{1}{r_k-\left(\frac{g-1}{g+1}\right)}\right]~=~\frac{i}{(g+1)}~\left[-\frac{1}{\left(\frac{g-1}{g+1}\right)^n-1}+\frac{\left(\frac{g-1}{g+1}\right)^{n-1}}{\left(\frac{g-1}{g+1}\right)^n-1}\right]\\
		&=i\left[\frac{(g+1)^{n-1}-(g-1)^{n-1}}{(g+1)^{n}-(g-1)^{n}}\right]=v_0(g).
	\end{split}
\end{equation}
Hence , for the entire matrix $\mathcal{G}^{ss}$,
\begin{equation}
	V_{\alpha\alpha}(\mathcal{G}^{ss})=i\left[\frac{(\mathcal{G}^{ss}+1)^{n-1}-(\mathcal{G}^{ss}-1)^{n-1}}{(\mathcal{G}^{ss}+1)^{n}-(\mathcal{G}^{ss}-1)^{n}}\right]=v_0(\mathcal{G}^{ss})
\end{equation}
For $\alpha>\beta$, the matrix element $\mathcal{V}_{\alpha\beta}$ becomes
\begin{equation}
	\begin{split}
		V_{\alpha\beta}(g)&=~\frac{i}{(g+1)n}~\sum_{k=1}^{n}~\left[\frac{r_k^{\alpha-\beta+1}}{r_k-\left(\frac{g-1}{g+1}\right)}-\frac{r_k^{\alpha-\beta}}{r_k-\left(\frac{g-1}{g+1}\right)}\right]~=~\frac{i}{(g+1)}~\left[-\frac{\left(\frac{g-1}{g+1}\right)^{\alpha-\beta}}{\left(\frac{g-1}{g+1}\right)^n-1}+\frac{\left(\frac{g-1}{g+1}\right)^{\alpha-\beta-1}}{\left(\frac{g-1}{g+1}\right)^n-1}\right]\\
		&=-2i~\left[\frac{(g+1)^{n-(\alpha-\beta)-1}(g-1)^{(\alpha-\beta)-1}}{(g+1)^{n}-(g-1)^{n}}\right]~=~v_{\alpha-\beta}(g).	
	\end{split}
\end{equation}
Hence, for the entire matrix $\mathcal{G}^{ss}$,
\begin{equation}
	V_{\alpha\beta}(\mathcal{G}^{ss})=~-2i~\left[\frac{(\mathcal{G}^{ss}+1)^{n-(\alpha-\beta)-1}(\mathcal{G}^{ss}-1)^{(\alpha-\beta)-1}}{(\mathcal{G}^{ss}+1)^{n}-(\mathcal{G}^{ss}-1)^{n}}\right]=v_{\alpha-\beta}(\mathcal{G}^{ss}).
\end{equation}
Similarly, the matrix element $\mathcal{V}_{\alpha\beta}$ for $\alpha<\beta$ becomes $v_{n-(\beta-\alpha)}$, which can be obtained by the transformation $(\alpha-\beta)\to n-(\beta-\alpha)$ or by a direct calculation. Hence, the final answer for the matrix elements of $\mathcal{V}$ can be written as $\mathcal{V}_{\alpha\beta}=\Theta(\alpha-\beta)v_{\alpha-\beta}+\Theta{(\beta-\alpha)}v_{n-(\beta-\alpha)}+\delta_{\alpha\beta}v_0$, which is the form shown in the main text. Note that $v_0$ has a different structure from the other matrix elements because it receives an additional contribution from a pole at $p=0$, unlike the others.

\section{Simplification of general results to Ground states:}\label{Ground_state_case}
According to Eq.~\eqref{Greensfn} in the main text , the equal time correlator matrix  $\mathcal{G}^{ss}$ at $t_0$ inside the subsystem $A$ is defined as
\begin{equation}
	i\mathcal{G}^{ss}(\rr,t_0;\rr',t_0) = \left[\begin{array}{cc} \langle \pi_s(\rr,t_0)\phi_s(\rr',t_0)\rangle & \langle \pi_s(\rr,t_0)\pi_s(\rr',t_0)\rangle \\ -\langle \phi_s(\rr,t_0)\phi_s(\rr',t_0)\rangle  & -\langle \phi_s(\rr,t_0)\pi_s(\rr',t_0)\rangle\end{array} \right]. 
\end{equation}
For thermal equilibrium, the equal time correlators $\langle\pi_s(\rr,t_0)\phi_s(\rr',t_0)\rangle$ and $\langle \phi_s(\rr,t_0)\pi_s(\rr',t_0)\rangle$ vanish whereas $\langle\phi_s(\rr,t_0)\phi_s(\rr',t_0)\rangle$ and $\langle \pi_s(\rr,t_0)\pi_s(\rr',t_0)\rangle$ remain finite. Let us define  $\langle\phi_s(\rr,t_0)\phi_s(\rr',t_0)\rangle=M_{+}(\rr,\rr')$ and $\langle\pi_s(\rr,t_0)\pi_s(\rr',t_0)\rangle=M_{-}(\rr,\rr')$ where the spatial indices $\rr,\rr'\in A$ i.e., run only over subsystem $A$. Then
\begin{equation}
	\begin{split}
		\mathcal{G}^{ss}=\left[\begin{array}{cc} 0 & -iM_{-}\\
			iM_{+} & 0\end{array}\right]; \text{~and~}(\mathcal{G}^{ss})^{2k}=\left[\begin{array}{cc} M_1^{2k}& 0\\
			0 & M_2^{2k}\end{array}\right];\\
	\end{split}
\end{equation}	
where we have defined $M_1^2=M_{-}M_{+}$ and $M_2^2=M_{+}M_{-}$. Then, for odd $n$,
\begin{equation}\label{g_plus1_power_n_minus_g_minus_1_power_n}
	\begin{split}
		(\mathcal{G}^{ss}+1)^n-(\mathcal{G}^{ss}-1)^n &=2\sum_{k=0}^{\frac{n-1}{2}}~^nC_{2k} (\mathcal{G}^{ss})^{2k}~=~\left[\begin{array}{cc} (M_1+1)^n-(M_1-1)^n &0
			\\0 & (M_2+1)^n-(M_2-1)^n\end{array}\right]\\
	\end{split}
\end{equation}
and hence
\begin{equation}
	\begin{split}
		\operatorname{Tr}\ln[(\mathcal{G}^{ss}+1)^n-(\mathcal{G}^{ss}-1)^n]&=\sum_{a=1,2}\operatorname{Tr}_A\ln[(M_a+1)^n-(M_a-1)^n]=2 \operatorname{Tr}_A\ln[(M+1)^n-(M-1)^n]
	\end{split}
\end{equation}
where $M^2=M_2^2=M_{+}M_{-}$, and we have used $\operatorname{Tr}[(M_-M_+)^k]=\operatorname{Tr}[(M_+M_-)^k]$. 

\section{Second order Feynman diagrams for $S^{(n)}$ in a $\phi^4$ theory}\label{Second_order_diagrams}

\begin{figure*}[t] \includegraphics[width=\textwidth]{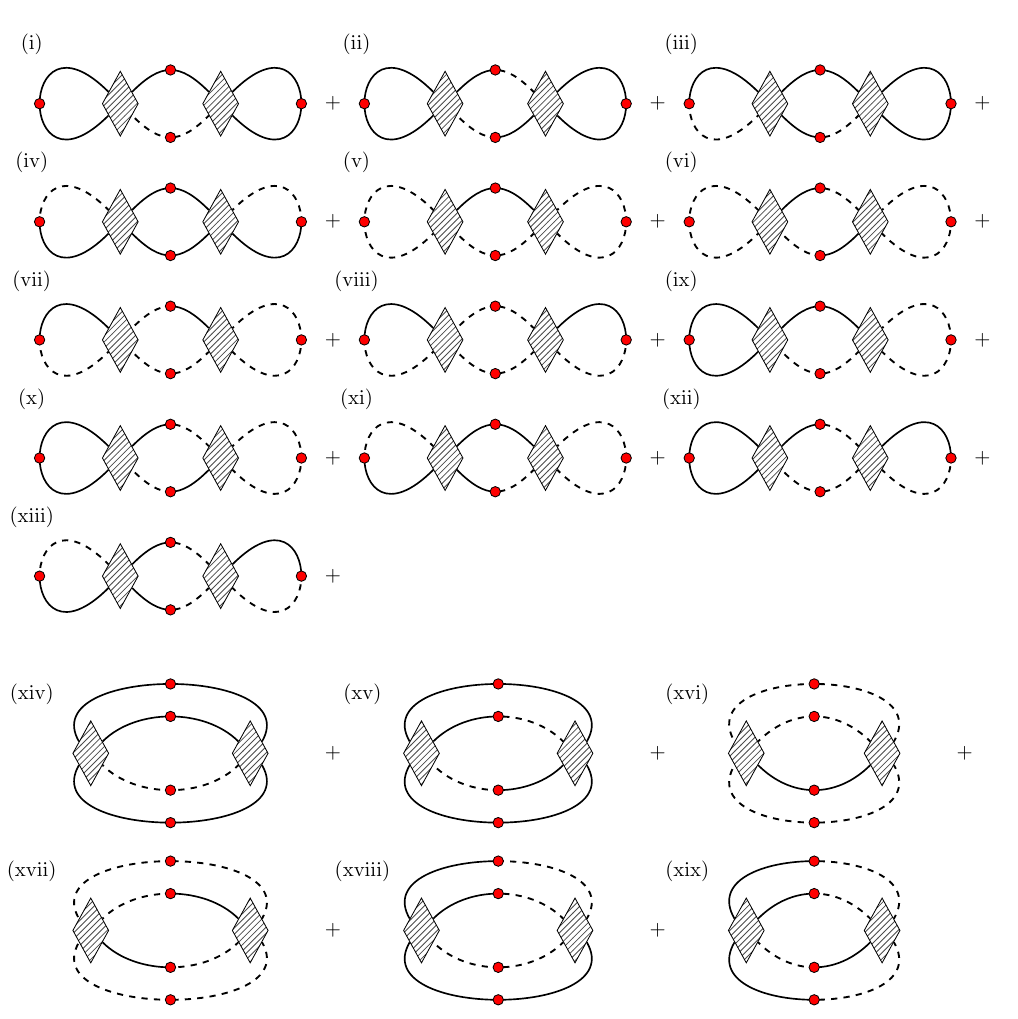}
	\caption{Second-order free-energy diagrams (i) - (xix) representing all connected ${\cal O}(\lambda^2)$ contributions to $S^{(n)}$. Out of all the nineteen diagrams, (v) and (vi) vanish in the $n\to 1$ limit as they contain two factors of the replica-diagonal propagators $\tilde{G}^{aa}_{\alpha\alpha}$.}
	\label{Second_order_Free_energy_Diagrams}
\end{figure*}

In the main text, we have demonstrated how to draw Feynman diagrams to calculate interaction corrections to $S^{(n)}$ in a $\phi^4$ theory. We have presented the first-order diagrams and the corresponding integrals in the main text. Here, we present all the Feynman diagrams for the $2^{\mathrm{nd}}$ order corrections in Fig.~\ref{Second_order_Free_energy_Diagrams}. The propagators with red circles are given by $\tilde{G}$, while the diamonds indicate the $\phi^4$ coupling constant. At this order, there are nineteen connected diagrams labelled (i)-(xix) in Fig.~\Ref{Second_order_Free_energy_Diagrams}. Each diagram can be evaluated using the Feynman rules provided in the main text. For example, the diagram shown in  Fig.~\Ref{Second_order_Free_energy_Diagrams}(i) evaluates to 
\begin{equation}
	\begin{split}
		-\frac{\lambda^2}{16} \sum_{\alpha,\beta=1}^{n}\int d^d \rr_1 \int dt_1 \int d^d \rr_2 \int dt_2& i\tilde{G}^{ss}_{\alpha\alpha}(\rr_1,t_1;\rr_1,t_1)\frac{1}{2}\left[i\tilde{G}^{ss}_{\alpha\beta}(\rr_1,t_1;\rr_2,t_2)+i\tilde{G}^{ss}_{\beta\alpha}(\rr_2,t_2;\rr_1,t_1)\right]\\
		&\times\frac{1}{2}\left[i\tilde{G}^{aa}_{\alpha\beta}(\rr_1,t_1;\rr_2,t_2)+i\tilde{G}^{aa}_{\beta\alpha}(\rr_2,t_2;\rr_1,t_1)\right]i\tilde{G}^{ss}_{\beta\beta}(\rr_2,t_2;\rr_2,t_2).
	\end{split}
\end{equation}

Analogous to the first order case where one diagram vanishes in the $n\to 1$ limit, the ${\cal O}(\lambda^2)$ diagrams labelled by (v) and (vi) also vanish explicitly when evaluating $S^{\mathrm{vN}}$.
Both diagrams scale as $\tilde{G}^{aa}_{\alpha\alpha}\tilde{G}^{aa}_{\beta\beta}\sim(n-1)^2$. Hence, after dividing it by $(n-1)$, their contribution to $S^{\mathrm{vN}}$ vanishes.

	\newpage
	\twocolumngrid
	\bibliography{Entanglement.bib}		

\begin{thebibliography}{74}%
\makeatletter
\providecommand \@ifxundefined [1]{%
 \@ifx{#1\undefined}
}%
\providecommand \@ifnum [1]{%
 \ifnum #1\expandafter \@firstoftwo
 \else \expandafter \@secondoftwo
 \fi
}%
\providecommand \@ifx [1]{%
 \ifx #1\expandafter \@firstoftwo
 \else \expandafter \@secondoftwo
 \fi
}%
\providecommand \natexlab [1]{#1}%
\providecommand \enquote  [1]{``#1''}%
\providecommand \bibnamefont  [1]{#1}%
\providecommand \bibfnamefont [1]{#1}%
\providecommand \citenamefont [1]{#1}%
\providecommand \href@noop [0]{\@secondoftwo}%
\providecommand \href [0]{\begingroup \@sanitize@url \@href}%
\providecommand \@href[1]{\@@startlink{#1}\@@href}%
\providecommand \@@href[1]{\endgroup#1\@@endlink}%
\providecommand \@sanitize@url [0]{\catcode `\\12\catcode `\$12\catcode
  `\&12\catcode `\#12\catcode `\^12\catcode `\_12\catcode `\%12\relax}%
\providecommand \@@startlink[1]{}%
\providecommand \@@endlink[0]{}%
\providecommand \url  [0]{\begingroup\@sanitize@url \@url }%
\providecommand \@url [1]{\endgroup\@href {#1}{\urlprefix }}%
\providecommand \urlprefix  [0]{URL }%
\providecommand \Eprint [0]{\href }%
\providecommand \doibase [0]{https://doi.org/}%
\providecommand \selectlanguage [0]{\@gobble}%
\providecommand \bibinfo  [0]{\@secondoftwo}%
\providecommand \bibfield  [0]{\@secondoftwo}%
\providecommand \translation [1]{[#1]}%
\providecommand \BibitemOpen [0]{}%
\providecommand \bibitemStop [0]{}%
\providecommand \bibitemNoStop [0]{.\EOS\space}%
\providecommand \EOS [0]{\spacefactor3000\relax}%
\providecommand \BibitemShut  [1]{\csname bibitem#1\endcsname}%
\let\auto@bib@innerbib\@empty
\bibitem [{\citenamefont {Schrödinger}(1935)}]{Schrodinger_1935}%
  \BibitemOpen
  \bibfield  {author} {\bibinfo {author} {\bibfnamefont {E.}~\bibnamefont
  {Schrödinger}},\ }\bibfield  {title} {\bibinfo {title} {Discussion of
  probability relations between separated systems},\ }\href
  {https://doi.org/10.1017/S0305004100013554} {\bibfield  {journal} {\bibinfo
  {journal} {Mathematical Proceedings of the Cambridge Philosophical Society}\
  }\textbf {\bibinfo {volume} {31}},\ \bibinfo {pages} {555–563} (\bibinfo
  {year} {1935})}\BibitemShut {NoStop}%
\bibitem [{\citenamefont {Schrödinger}(1936)}]{Schrodinger_1936}%
  \BibitemOpen
  \bibfield  {author} {\bibinfo {author} {\bibfnamefont {E.}~\bibnamefont
  {Schrödinger}},\ }\bibfield  {title} {\bibinfo {title} {Probability
  relations between separated systems},\ }\href
  {https://doi.org/10.1017/S0305004100019137} {\bibfield  {journal} {\bibinfo
  {journal} {Mathematical Proceedings of the Cambridge Philosophical Society}\
  }\textbf {\bibinfo {volume} {32}},\ \bibinfo {pages} {446–452} (\bibinfo
  {year} {1936})}\BibitemShut {NoStop}%
\bibitem [{\citenamefont {Einstein}\ \emph {et~al.}(1935)\citenamefont
  {Einstein}, \citenamefont {Podolsky},\ and\ \citenamefont
  {Rosen}}]{EPR_main_paper}%
  \BibitemOpen
  \bibfield  {author} {\bibinfo {author} {\bibfnamefont {A.}~\bibnamefont
  {Einstein}}, \bibinfo {author} {\bibfnamefont {B.}~\bibnamefont {Podolsky}},\
  and\ \bibinfo {author} {\bibfnamefont {N.}~\bibnamefont {Rosen}},\ }\bibfield
   {title} {\bibinfo {title} {Can quantum-mechanical description of physical
  reality be considered complete?},\ }\href
  {https://doi.org/10.1103/PhysRev.47.777} {\bibfield  {journal} {\bibinfo
  {journal} {Phys. Rev.}\ }\textbf {\bibinfo {volume} {47}},\ \bibinfo {pages}
  {777} (\bibinfo {year} {1935})}\BibitemShut {NoStop}%
\bibitem [{\citenamefont {Bohm}\ and\ \citenamefont
  {Aharonov}(1957)}]{BohmEPR}%
  \BibitemOpen
  \bibfield  {author} {\bibinfo {author} {\bibfnamefont {D.}~\bibnamefont
  {Bohm}}\ and\ \bibinfo {author} {\bibfnamefont {Y.}~\bibnamefont
  {Aharonov}},\ }\bibfield  {title} {\bibinfo {title} {Discussion of
  experimental proof for the paradox of einstein, rosen, and podolsky},\ }\href
  {https://doi.org/10.1103/PhysRev.108.1070} {\bibfield  {journal} {\bibinfo
  {journal} {Phys. Rev.}\ }\textbf {\bibinfo {volume} {108}},\ \bibinfo {pages}
  {1070} (\bibinfo {year} {1957})}\BibitemShut {NoStop}%
\bibitem [{\citenamefont {Bell}(1964)}]{Bells_inequality}%
  \BibitemOpen
  \bibfield  {author} {\bibinfo {author} {\bibfnamefont {J.~S.}\ \bibnamefont
  {Bell}},\ }\bibfield  {title} {\bibinfo {title} {On the einstein podolsky
  rosen paradox},\ }\href {https://doi.org/10.1103/PhysicsPhysiqueFizika.1.195}
  {\bibfield  {journal} {\bibinfo  {journal} {Physics Physique Fizika}\
  }\textbf {\bibinfo {volume} {1}},\ \bibinfo {pages} {195} (\bibinfo {year}
  {1964})}\BibitemShut {NoStop}%
\bibitem [{\citenamefont {Clauser}\ \emph {et~al.}(1969)\citenamefont
  {Clauser}, \citenamefont {Horne}, \citenamefont {Shimony},\ and\
  \citenamefont {Holt}}]{CHSH}%
  \BibitemOpen
  \bibfield  {author} {\bibinfo {author} {\bibfnamefont {J.~F.}\ \bibnamefont
  {Clauser}}, \bibinfo {author} {\bibfnamefont {M.~A.}\ \bibnamefont {Horne}},
  \bibinfo {author} {\bibfnamefont {A.}~\bibnamefont {Shimony}},\ and\ \bibinfo
  {author} {\bibfnamefont {R.~A.}\ \bibnamefont {Holt}},\ }\bibfield  {title}
  {\bibinfo {title} {Proposed experiment to test local hidden-variable
  theories},\ }\href {https://doi.org/10.1103/PhysRevLett.23.880} {\bibfield
  {journal} {\bibinfo  {journal} {Phys. Rev. Lett.}\ }\textbf {\bibinfo
  {volume} {23}},\ \bibinfo {pages} {880} (\bibinfo {year} {1969})}\BibitemShut
  {NoStop}%
\bibitem [{\citenamefont {Aspect}\ \emph
  {et~al.}(1982{\natexlab{a}})\citenamefont {Aspect}, \citenamefont
  {Dalibard},\ and\ \citenamefont {Roger}}]{Aspect_1}%
  \BibitemOpen
  \bibfield  {author} {\bibinfo {author} {\bibfnamefont {A.}~\bibnamefont
  {Aspect}}, \bibinfo {author} {\bibfnamefont {J.}~\bibnamefont {Dalibard}},\
  and\ \bibinfo {author} {\bibfnamefont {G.}~\bibnamefont {Roger}},\ }\bibfield
   {title} {\bibinfo {title} {Experimental test of bell's inequalities using
  time-varying analyzers},\ }\href
  {https://doi.org/10.1103/PhysRevLett.49.1804} {\bibfield  {journal} {\bibinfo
   {journal} {Phys. Rev. Lett.}\ }\textbf {\bibinfo {volume} {49}},\ \bibinfo
  {pages} {1804} (\bibinfo {year} {1982}{\natexlab{a}})}\BibitemShut {NoStop}%
\bibitem [{\citenamefont {Aspect}\ \emph
  {et~al.}(1982{\natexlab{b}})\citenamefont {Aspect}, \citenamefont
  {Grangier},\ and\ \citenamefont {Roger}}]{Aspect_2}%
  \BibitemOpen
  \bibfield  {author} {\bibinfo {author} {\bibfnamefont {A.}~\bibnamefont
  {Aspect}}, \bibinfo {author} {\bibfnamefont {P.}~\bibnamefont {Grangier}},\
  and\ \bibinfo {author} {\bibfnamefont {G.}~\bibnamefont {Roger}},\ }\bibfield
   {title} {\bibinfo {title} {Experimental realization of
  einstein-podolsky-rosen-bohm gedankenexperiment: A new violation of bell's
  inequalities},\ }\href {https://doi.org/10.1103/PhysRevLett.49.91} {\bibfield
   {journal} {\bibinfo  {journal} {Phys. Rev. Lett.}\ }\textbf {\bibinfo
  {volume} {49}},\ \bibinfo {pages} {91} (\bibinfo {year}
  {1982}{\natexlab{b}})}\BibitemShut {NoStop}%
\bibitem [{\citenamefont {Zeilinger}(1999)}]{Zeilinger_expt}%
  \BibitemOpen
  \bibfield  {author} {\bibinfo {author} {\bibfnamefont {A.}~\bibnamefont
  {Zeilinger}},\ }\bibfield  {title} {\bibinfo {title} {Experiment and the
  foundations of quantum physics},\ }\href
  {https://doi.org/10.1103/RevModPhys.71.S288} {\bibfield  {journal} {\bibinfo
  {journal} {Rev. Mod. Phys.}\ }\textbf {\bibinfo {volume} {71}},\ \bibinfo
  {pages} {S288} (\bibinfo {year} {1999})}\BibitemShut {NoStop}%
\bibitem [{\citenamefont {Laflorencie}(2016)}]{LAFLORENCIE20161}%
  \BibitemOpen
  \bibfield  {author} {\bibinfo {author} {\bibfnamefont {N.}~\bibnamefont
  {Laflorencie}},\ }\bibfield  {title} {\bibinfo {title} {Quantum entanglement
  in condensed matter systems},\ }\href
  {https://doi.org/https://doi.org/10.1016/j.physrep.2016.06.008} {\bibfield
  {journal} {\bibinfo  {journal} {Physics Reports}\ }\textbf {\bibinfo {volume}
  {646}},\ \bibinfo {pages} {1} (\bibinfo {year} {2016})}\BibitemShut {NoStop}%
\bibitem [{\citenamefont {Jiang}\ \emph {et~al.}(2012)\citenamefont {Jiang},
  \citenamefont {Wang},\ and\ \citenamefont {Balents}}]{Jiang2012}%
  \BibitemOpen
  \bibfield  {author} {\bibinfo {author} {\bibfnamefont {H.-C.}\ \bibnamefont
  {Jiang}}, \bibinfo {author} {\bibfnamefont {Z.}~\bibnamefont {Wang}},\ and\
  \bibinfo {author} {\bibfnamefont {L.}~\bibnamefont {Balents}},\ }\bibfield
  {title} {\bibinfo {title} {Identifying topological order by entanglement
  entropy},\ }\href {https://doi.org/10.1038/nphys2465} {\bibfield  {journal}
  {\bibinfo  {journal} {Nature Physics}\ }\textbf {\bibinfo {volume} {8}},\
  \bibinfo {pages} {902} (\bibinfo {year} {2012})}\BibitemShut {NoStop}%
\bibitem [{\citenamefont {Levin}\ and\ \citenamefont
  {Wen}(2006)}]{LevinWen_Topo}%
  \BibitemOpen
  \bibfield  {author} {\bibinfo {author} {\bibfnamefont {M.}~\bibnamefont
  {Levin}}\ and\ \bibinfo {author} {\bibfnamefont {X.-G.}\ \bibnamefont
  {Wen}},\ }\bibfield  {title} {\bibinfo {title} {Detecting topological order
  in a ground state wave function},\ }\href
  {https://doi.org/10.1103/PhysRevLett.96.110405} {\bibfield  {journal}
  {\bibinfo  {journal} {Phys. Rev. Lett.}\ }\textbf {\bibinfo {volume} {96}},\
  \bibinfo {pages} {110405} (\bibinfo {year} {2006})}\BibitemShut {NoStop}%
\bibitem [{\citenamefont {Isakov}\ \emph {et~al.}(2011)\citenamefont {Isakov},
  \citenamefont {Hastings},\ and\ \citenamefont {Melko}}]{HastingsMelko}%
  \BibitemOpen
  \bibfield  {author} {\bibinfo {author} {\bibfnamefont {S.~V.}\ \bibnamefont
  {Isakov}}, \bibinfo {author} {\bibfnamefont {M.~B.}\ \bibnamefont
  {Hastings}},\ and\ \bibinfo {author} {\bibfnamefont {R.~G.}\ \bibnamefont
  {Melko}},\ }\bibfield  {title} {\bibinfo {title} {Topological entanglement
  entropy of a bose--hubbard spin liquid},\ }\href
  {https://doi.org/10.1038/nphys2036} {\bibfield  {journal} {\bibinfo
  {journal} {Nature Physics}\ }\textbf {\bibinfo {volume} {7}},\ \bibinfo
  {pages} {772} (\bibinfo {year} {2011})}\BibitemShut {NoStop}%
\bibitem [{\citenamefont {Hertzberg}\ and\ \citenamefont
  {Wilczek}(2011)}]{Wilczek}%
  \BibitemOpen
  \bibfield  {author} {\bibinfo {author} {\bibfnamefont {M.~P.}\ \bibnamefont
  {Hertzberg}}\ and\ \bibinfo {author} {\bibfnamefont {F.}~\bibnamefont
  {Wilczek}},\ }\bibfield  {title} {\bibinfo {title} {Some calculable
  contributions to entanglement entropy},\ }\href
  {https://doi.org/10.1103/PhysRevLett.106.050404} {\bibfield  {journal}
  {\bibinfo  {journal} {Phys. Rev. Lett.}\ }\textbf {\bibinfo {volume} {106}},\
  \bibinfo {pages} {050404} (\bibinfo {year} {2011})}\BibitemShut {NoStop}%
\bibitem [{\citenamefont {Calabrese}\ and\ \citenamefont
  {Cardy}(2009)}]{Calabrese_2009}%
  \BibitemOpen
  \bibfield  {author} {\bibinfo {author} {\bibfnamefont {P.}~\bibnamefont
  {Calabrese}}\ and\ \bibinfo {author} {\bibfnamefont {J.}~\bibnamefont
  {Cardy}},\ }\bibfield  {title} {\bibinfo {title} {Entanglement entropy and
  conformal field theory},\ }\href
  {https://doi.org/10.1088/1751-8113/42/50/504005} {\bibfield  {journal}
  {\bibinfo  {journal} {Journal of Physics A: Mathematical and Theoretical}\
  }\textbf {\bibinfo {volume} {42}},\ \bibinfo {pages} {504005} (\bibinfo
  {year} {2009})}\BibitemShut {NoStop}%
\bibitem [{\citenamefont {Calabrese}\ and\ \citenamefont
  {Cardy}(2004)}]{CalabreseCardy}%
  \BibitemOpen
  \bibfield  {author} {\bibinfo {author} {\bibfnamefont {P.}~\bibnamefont
  {Calabrese}}\ and\ \bibinfo {author} {\bibfnamefont {J.}~\bibnamefont
  {Cardy}},\ }\bibfield  {title} {\bibinfo {title} {Entanglement entropy and
  quantum field theory},\ }\href
  {https://doi.org/10.1088/1742-5468/2004/06/p06002} {\bibfield  {journal}
  {\bibinfo  {journal} {Journal of Statistical Mechanics: Theory and
  Experiment}\ }\textbf {\bibinfo {volume} {2004}},\ \bibinfo {pages} {P06002}
  (\bibinfo {year} {2004})}\BibitemShut {NoStop}%
\bibitem [{\citenamefont {Casini}\ and\ \citenamefont
  {Huerta}(2009)}]{Casini_2009}%
  \BibitemOpen
  \bibfield  {author} {\bibinfo {author} {\bibfnamefont {H.}~\bibnamefont
  {Casini}}\ and\ \bibinfo {author} {\bibfnamefont {M.}~\bibnamefont
  {Huerta}},\ }\bibfield  {title} {\bibinfo {title} {Entanglement entropy in
  free quantum field theory},\ }\href
  {https://doi.org/10.1088/1751-8113/42/50/504007} {\bibfield  {journal}
  {\bibinfo  {journal} {Journal of Physics A: Mathematical and Theoretical}\
  }\textbf {\bibinfo {volume} {42}},\ \bibinfo {pages} {504007} (\bibinfo
  {year} {2009})}\BibitemShut {NoStop}%
\bibitem [{\citenamefont {Ryu}\ and\ \citenamefont
  {Takayanagi}(2006)}]{RyuTakayanagi2006}%
  \BibitemOpen
  \bibfield  {author} {\bibinfo {author} {\bibfnamefont {S.}~\bibnamefont
  {Ryu}}\ and\ \bibinfo {author} {\bibfnamefont {T.}~\bibnamefont
  {Takayanagi}},\ }\bibfield  {title} {\bibinfo {title} {Aspects of holographic
  entanglement entropy},\ }\href
  {https://doi.org/10.1088/1126-6708/2006/08/045} {\bibfield  {journal}
  {\bibinfo  {journal} {Journal of High Energy Physics}\ }\textbf {\bibinfo
  {volume} {2006}},\ \bibinfo {pages} {045} (\bibinfo {year}
  {2006})}\BibitemShut {NoStop}%
\bibitem [{\citenamefont {Srednicki}(1994)}]{SrednickiETH}%
  \BibitemOpen
  \bibfield  {author} {\bibinfo {author} {\bibfnamefont {M.}~\bibnamefont
  {Srednicki}},\ }\bibfield  {title} {\bibinfo {title} {Chaos and quantum
  thermalization},\ }\href {https://doi.org/10.1103/PhysRevE.50.888} {\bibfield
   {journal} {\bibinfo  {journal} {Phys. Rev. E}\ }\textbf {\bibinfo {volume}
  {50}},\ \bibinfo {pages} {888} (\bibinfo {year} {1994})}\BibitemShut
  {NoStop}%
\bibitem [{\citenamefont {Deutsch}(2018)}]{Deutsch_2018}%
  \BibitemOpen
  \bibfield  {author} {\bibinfo {author} {\bibfnamefont {J.~M.}\ \bibnamefont
  {Deutsch}},\ }\bibfield  {title} {\bibinfo {title} {Eigenstate thermalization
  hypothesis},\ }\href {https://doi.org/10.1088/1361-6633/aac9f1} {\bibfield
  {journal} {\bibinfo  {journal} {Reports on Progress in Physics}\ }\textbf
  {\bibinfo {volume} {81}},\ \bibinfo {pages} {082001} (\bibinfo {year}
  {2018})}\BibitemShut {NoStop}%
\bibitem [{\citenamefont {Nandkishore}\ and\ \citenamefont
  {Huse}(2015)}]{Nandkishore}%
  \BibitemOpen
  \bibfield  {author} {\bibinfo {author} {\bibfnamefont {R.}~\bibnamefont
  {Nandkishore}}\ and\ \bibinfo {author} {\bibfnamefont {D.~A.}\ \bibnamefont
  {Huse}},\ }\bibfield  {title} {\bibinfo {title} {{Many body localization and
  thermalization in quantum statistical mechanics}},\ }\href
  {https://doi.org/10.1146/annurev-conmatphys-031214-014726} {\bibfield
  {journal} {\bibinfo  {journal} {Ann. Rev. Condensed Matter Phys.}\ }\textbf
  {\bibinfo {volume} {6}},\ \bibinfo {pages} {15} (\bibinfo {year} {2015})},\
  \Eprint {https://arxiv.org/abs/1404.0686} {arXiv:1404.0686
  [cond-mat.stat-mech]} \BibitemShut {NoStop}%
\bibitem [{\citenamefont {Pal}\ and\ \citenamefont {Huse}(2010)}]{HusePal}%
  \BibitemOpen
  \bibfield  {author} {\bibinfo {author} {\bibfnamefont {A.}~\bibnamefont
  {Pal}}\ and\ \bibinfo {author} {\bibfnamefont {D.~A.}\ \bibnamefont {Huse}},\
  }\bibfield  {title} {\bibinfo {title} {Many-body localization phase
  transition},\ }\href {https://doi.org/10.1103/PhysRevB.82.174411} {\bibfield
  {journal} {\bibinfo  {journal} {Phys. Rev. B}\ }\textbf {\bibinfo {volume}
  {82}},\ \bibinfo {pages} {174411} (\bibinfo {year} {2010})}\BibitemShut
  {NoStop}%
\bibitem [{\citenamefont {Abanin}\ \emph {et~al.}(2019)\citenamefont {Abanin},
  \citenamefont {Altman}, \citenamefont {Bloch},\ and\ \citenamefont
  {Serbyn}}]{abanin2019colloquium}%
  \BibitemOpen
  \bibfield  {author} {\bibinfo {author} {\bibfnamefont {D.~A.}\ \bibnamefont
  {Abanin}}, \bibinfo {author} {\bibfnamefont {E.}~\bibnamefont {Altman}},
  \bibinfo {author} {\bibfnamefont {I.}~\bibnamefont {Bloch}},\ and\ \bibinfo
  {author} {\bibfnamefont {M.}~\bibnamefont {Serbyn}},\ }\bibfield  {title}
  {\bibinfo {title} {Colloquium: Many-body localization, thermalization, and
  entanglement},\ }\href@noop {} {\bibfield  {journal} {\bibinfo  {journal}
  {Reviews of Modern Physics}\ }\textbf {\bibinfo {volume} {91}},\ \bibinfo
  {pages} {021001} (\bibinfo {year} {2019})}\BibitemShut {NoStop}%
\bibitem [{\citenamefont {Wang}\ \emph {et~al.}(2004)\citenamefont {Wang},
  \citenamefont {Ghose}, \citenamefont {Sanders},\ and\ \citenamefont
  {Hu}}]{HuPRE70_2004}%
  \BibitemOpen
  \bibfield  {author} {\bibinfo {author} {\bibfnamefont {X.}~\bibnamefont
  {Wang}}, \bibinfo {author} {\bibfnamefont {S.}~\bibnamefont {Ghose}},
  \bibinfo {author} {\bibfnamefont {B.~C.}\ \bibnamefont {Sanders}},\ and\
  \bibinfo {author} {\bibfnamefont {B.}~\bibnamefont {Hu}},\ }\bibfield
  {title} {\bibinfo {title} {Entanglement as a signature of quantum chaos},\
  }\href {https://doi.org/10.1103/PhysRevE.70.016217} {\bibfield  {journal}
  {\bibinfo  {journal} {Phys. Rev. E}\ }\textbf {\bibinfo {volume} {70}},\
  \bibinfo {pages} {016217} (\bibinfo {year} {2004})}\BibitemShut {NoStop}%
\bibitem [{\citenamefont {Mej\'{\i}a-Monasterio}\ \emph
  {et~al.}(2005)\citenamefont {Mej\'{\i}a-Monasterio}, \citenamefont {Benenti},
  \citenamefont {Carlo},\ and\ \citenamefont {Casati}}]{CasatiPRA71_2005}%
  \BibitemOpen
  \bibfield  {author} {\bibinfo {author} {\bibfnamefont {C.}~\bibnamefont
  {Mej\'{\i}a-Monasterio}}, \bibinfo {author} {\bibfnamefont {G.}~\bibnamefont
  {Benenti}}, \bibinfo {author} {\bibfnamefont {G.~G.}\ \bibnamefont {Carlo}},\
  and\ \bibinfo {author} {\bibfnamefont {G.}~\bibnamefont {Casati}},\
  }\bibfield  {title} {\bibinfo {title} {Entanglement across a transition to
  quantum chaos},\ }\href {https://doi.org/10.1103/PhysRevA.71.062324}
  {\bibfield  {journal} {\bibinfo  {journal} {Phys. Rev. A}\ }\textbf {\bibinfo
  {volume} {71}},\ \bibinfo {pages} {062324} (\bibinfo {year}
  {2005})}\BibitemShut {NoStop}%
\bibitem [{\citenamefont {Lantagne-Hurtubise}\ \emph
  {et~al.}(2020)\citenamefont {Lantagne-Hurtubise}, \citenamefont {Plugge},
  \citenamefont {Can},\ and\ \citenamefont {Franz}}]{Franz_2_2020}%
  \BibitemOpen
  \bibfield  {author} {\bibinfo {author} {\bibfnamefont {E.}~\bibnamefont
  {Lantagne-Hurtubise}}, \bibinfo {author} {\bibfnamefont {S.}~\bibnamefont
  {Plugge}}, \bibinfo {author} {\bibfnamefont {O.}~\bibnamefont {Can}},\ and\
  \bibinfo {author} {\bibfnamefont {M.}~\bibnamefont {Franz}},\ }\bibfield
  {title} {\bibinfo {title} {Diagnosing quantum chaos in many-body systems
  using entanglement as a resource},\ }\href
  {https://doi.org/10.1103/PhysRevResearch.2.013254} {\bibfield  {journal}
  {\bibinfo  {journal} {Phys. Rev. Res.}\ }\textbf {\bibinfo {volume} {2}},\
  \bibinfo {pages} {013254} (\bibinfo {year} {2020})}\BibitemShut {NoStop}%
\bibitem [{\citenamefont {Gong}\ \emph {et~al.}(2021)\citenamefont {Gong},
  \citenamefont {Piroli},\ and\ \citenamefont {Cirac}}]{CiracPRL126_2021}%
  \BibitemOpen
  \bibfield  {author} {\bibinfo {author} {\bibfnamefont {Z.}~\bibnamefont
  {Gong}}, \bibinfo {author} {\bibfnamefont {L.}~\bibnamefont {Piroli}},\ and\
  \bibinfo {author} {\bibfnamefont {J.~I.}\ \bibnamefont {Cirac}},\ }\bibfield
  {title} {\bibinfo {title} {Topological lower bound on quantum chaos by
  entanglement growth},\ }\href
  {https://doi.org/10.1103/PhysRevLett.126.160601} {\bibfield  {journal}
  {\bibinfo  {journal} {Phys. Rev. Lett.}\ }\textbf {\bibinfo {volume} {126}},\
  \bibinfo {pages} {160601} (\bibinfo {year} {2021})}\BibitemShut {NoStop}%
\bibitem [{\citenamefont {Preskill}(2018)}]{preskill2018quantum}%
  \BibitemOpen
  \bibfield  {author} {\bibinfo {author} {\bibfnamefont {J.}~\bibnamefont
  {Preskill}},\ }\bibfield  {title} {\bibinfo {title} {Quantum computing in the
  nisq era and beyond},\ }\href@noop {} {\bibfield  {journal} {\bibinfo
  {journal} {Quantum}\ }\textbf {\bibinfo {volume} {2}},\ \bibinfo {pages} {79}
  (\bibinfo {year} {2018})}\BibitemShut {NoStop}%
\bibitem [{\citenamefont
  {Gottesman}(1998)}]{gottesman1998heisenbergrepresentationquantumcomputers}%
  \BibitemOpen
  \bibfield  {author} {\bibinfo {author} {\bibfnamefont {D.}~\bibnamefont
  {Gottesman}},\ }\href {https://arxiv.org/abs/quant-ph/9807006} {\bibinfo
  {title} {The heisenberg representation of quantum computers}} (\bibinfo
  {year} {1998}),\ \Eprint {https://arxiv.org/abs/quant-ph/9807006}
  {arXiv:quant-ph/9807006 [quant-ph]} \BibitemShut {NoStop}%
\bibitem [{\citenamefont {Islam}\ \emph {et~al.}(2015)\citenamefont {Islam},
  \citenamefont {Ma}, \citenamefont {Preiss}, \citenamefont {Eric~Tai},
  \citenamefont {Lukin}, \citenamefont {Rispoli},\ and\ \citenamefont
  {Greiner}}]{Islam2015_expt}%
  \BibitemOpen
  \bibfield  {author} {\bibinfo {author} {\bibfnamefont {R.}~\bibnamefont
  {Islam}}, \bibinfo {author} {\bibfnamefont {R.}~\bibnamefont {Ma}}, \bibinfo
  {author} {\bibfnamefont {P.~M.}\ \bibnamefont {Preiss}}, \bibinfo {author}
  {\bibfnamefont {M.}~\bibnamefont {Eric~Tai}}, \bibinfo {author}
  {\bibfnamefont {A.}~\bibnamefont {Lukin}}, \bibinfo {author} {\bibfnamefont
  {M.}~\bibnamefont {Rispoli}},\ and\ \bibinfo {author} {\bibfnamefont
  {M.}~\bibnamefont {Greiner}},\ }\bibfield  {title} {\bibinfo {title}
  {Measuring entanglement entropy in a quantum many-body system},\ }\href
  {https://doi.org/10.1038/nature15750} {\bibfield  {journal} {\bibinfo
  {journal} {Nature}\ }\textbf {\bibinfo {volume} {528}},\ \bibinfo {pages} {77
  EP } (\bibinfo {year} {2015})}\BibitemShut {NoStop}%
\bibitem [{\citenamefont {Schreiber}\ \emph {et~al.}(2015)\citenamefont
  {Schreiber}, \citenamefont {Hodgman}, \citenamefont {Bordia}, \citenamefont
  {L{\"u}schen}, \citenamefont {Fischer}, \citenamefont {Vosk}, \citenamefont
  {Altman}, \citenamefont {Schneider},\ and\ \citenamefont
  {Bloch}}]{Schreiber_MBL}%
  \BibitemOpen
  \bibfield  {author} {\bibinfo {author} {\bibfnamefont {M.}~\bibnamefont
  {Schreiber}}, \bibinfo {author} {\bibfnamefont {S.~S.}\ \bibnamefont
  {Hodgman}}, \bibinfo {author} {\bibfnamefont {P.}~\bibnamefont {Bordia}},
  \bibinfo {author} {\bibfnamefont {H.~P.}\ \bibnamefont {L{\"u}schen}},
  \bibinfo {author} {\bibfnamefont {M.~H.}\ \bibnamefont {Fischer}}, \bibinfo
  {author} {\bibfnamefont {R.}~\bibnamefont {Vosk}}, \bibinfo {author}
  {\bibfnamefont {E.}~\bibnamefont {Altman}}, \bibinfo {author} {\bibfnamefont
  {U.}~\bibnamefont {Schneider}},\ and\ \bibinfo {author} {\bibfnamefont
  {I.}~\bibnamefont {Bloch}},\ }\bibfield  {title} {\bibinfo {title}
  {Observation of many-body localization of interacting fermions in a
  quasirandom optical lattice},\ }\href
  {https://doi.org/10.1126/science.aaa7432} {\bibfield  {journal} {\bibinfo
  {journal} {Science}\ }\textbf {\bibinfo {volume} {349}},\ \bibinfo {pages}
  {842} (\bibinfo {year} {2015})}\BibitemShut {NoStop}%
\bibitem [{\citenamefont {Tajik}\ \emph {et~al.}(2023)\citenamefont {Tajik},
  \citenamefont {Kukuljan}, \citenamefont {Sotiriadis}, \citenamefont {Rauer},
  \citenamefont {Schweigler}, \citenamefont {Cataldini}, \citenamefont
  {Sabino}, \citenamefont {M{\o}ller}, \citenamefont {Sch{\"u}ttelkopf},
  \citenamefont {Ji}, \citenamefont {Sels}, \citenamefont {Demler},\ and\
  \citenamefont {Schmiedmayer}}]{Tajik2023}%
  \BibitemOpen
  \bibfield  {author} {\bibinfo {author} {\bibfnamefont {M.}~\bibnamefont
  {Tajik}}, \bibinfo {author} {\bibfnamefont {I.}~\bibnamefont {Kukuljan}},
  \bibinfo {author} {\bibfnamefont {S.}~\bibnamefont {Sotiriadis}}, \bibinfo
  {author} {\bibfnamefont {B.}~\bibnamefont {Rauer}}, \bibinfo {author}
  {\bibfnamefont {T.}~\bibnamefont {Schweigler}}, \bibinfo {author}
  {\bibfnamefont {F.}~\bibnamefont {Cataldini}}, \bibinfo {author}
  {\bibfnamefont {J.}~\bibnamefont {Sabino}}, \bibinfo {author} {\bibfnamefont
  {F.}~\bibnamefont {M{\o}ller}}, \bibinfo {author} {\bibfnamefont
  {P.}~\bibnamefont {Sch{\"u}ttelkopf}}, \bibinfo {author} {\bibfnamefont
  {S.-C.}\ \bibnamefont {Ji}}, \bibinfo {author} {\bibfnamefont
  {D.}~\bibnamefont {Sels}}, \bibinfo {author} {\bibfnamefont {E.}~\bibnamefont
  {Demler}},\ and\ \bibinfo {author} {\bibfnamefont {J.}~\bibnamefont
  {Schmiedmayer}},\ }\bibfield  {title} {\bibinfo {title} {Verification of the
  area law of mutual information in a quantum field simulator},\ }\href
  {https://doi.org/10.1038/s41567-023-02027-1} {\bibfield  {journal} {\bibinfo
  {journal} {Nature Physics}\ }\textbf {\bibinfo {volume} {19}},\ \bibinfo
  {pages} {1022} (\bibinfo {year} {2023})}\BibitemShut {NoStop}%
\bibitem [{\citenamefont {Srednicki}(2007)}]{Srednicki_2007_book}%
  \BibitemOpen
  \bibfield  {author} {\bibinfo {author} {\bibfnamefont {M.}~\bibnamefont
  {Srednicki}},\ }\href@noop {} {\emph {\bibinfo {title} {Quantum Field
  Theory}}}\ (\bibinfo  {publisher} {Cambridge University Press},\ \bibinfo
  {year} {2007})\BibitemShut {NoStop}%
\bibitem [{\citenamefont {Peskin}\ and\ \citenamefont
  {Schroeder}(1995)}]{Peskin_1995ev}%
  \BibitemOpen
  \bibfield  {author} {\bibinfo {author} {\bibfnamefont {M.~E.}\ \bibnamefont
  {Peskin}}\ and\ \bibinfo {author} {\bibfnamefont {D.~V.}\ \bibnamefont
  {Schroeder}},\ }\href {https://doi.org/10.1201/9780429503559} {\emph
  {\bibinfo {title} {{An Introduction to quantum field theory}}}}\ (\bibinfo
  {publisher} {Addison-Wesley},\ \bibinfo {address} {Reading, USA},\ \bibinfo
  {year} {1995})\BibitemShut {NoStop}%
\bibitem [{\citenamefont {Bassett}\ \emph {et~al.}(2006)\citenamefont
  {Bassett}, \citenamefont {Tsujikawa},\ and\ \citenamefont
  {Wands}}]{Bassett2006}%
  \BibitemOpen
  \bibfield  {author} {\bibinfo {author} {\bibfnamefont {B.~A.}\ \bibnamefont
  {Bassett}}, \bibinfo {author} {\bibfnamefont {S.}~\bibnamefont {Tsujikawa}},\
  and\ \bibinfo {author} {\bibfnamefont {D.}~\bibnamefont {Wands}},\ }\bibfield
   {title} {\bibinfo {title} {Inflation dynamics and reheating},\ }\href
  {https://doi.org/10.1103/RevModPhys.78.537} {\bibfield  {journal} {\bibinfo
  {journal} {Rev. Mod. Phys.}\ }\textbf {\bibinfo {volume} {78}},\ \bibinfo
  {pages} {537} (\bibinfo {year} {2006})}\BibitemShut {NoStop}%
\bibitem [{\citenamefont {Sachdev}(2011)}]{Sachdev_2011_Book}%
  \BibitemOpen
  \bibfield  {author} {\bibinfo {author} {\bibfnamefont {S.}~\bibnamefont
  {Sachdev}},\ }\href@noop {} {\emph {\bibinfo {title} {Quantum Phase
  Transitions}}},\ \bibinfo {edition} {2nd}\ ed.\ (\bibinfo  {publisher}
  {Cambridge University Press},\ \bibinfo {year} {2011})\BibitemShut {NoStop}%
\bibitem [{\citenamefont {Doniach}\ and\ \citenamefont
  {Sondheimer}(1998)}]{doniach}%
  \BibitemOpen
  \bibfield  {author} {\bibinfo {author} {\bibfnamefont {S.}~\bibnamefont
  {Doniach}}\ and\ \bibinfo {author} {\bibfnamefont {E.~H.}\ \bibnamefont
  {Sondheimer}},\ }\href {https://doi.org/10.1142/p067} {\emph {\bibinfo
  {title} {Green's Functions for Solid State Physicists}}}\ (\bibinfo
  {publisher} {Published by Imperial College press and distributed by World
  Scientific Publishing Co.},\ \bibinfo {year} {1998})\BibitemShut {NoStop}%
\bibitem [{\citenamefont {Patton}(1984)}]{magnon}%
  \BibitemOpen
  \bibfield  {author} {\bibinfo {author} {\bibfnamefont {C.~E.}\ \bibnamefont
  {Patton}},\ }\bibfield  {title} {\bibinfo {title} {Magnetic excitations in
  solids},\ }\href
  {https://doi.org/https://doi.org/10.1016/0370-1573(84)90023-1} {\bibfield
  {journal} {\bibinfo  {journal} {Physics Reports}\ }\textbf {\bibinfo {volume}
  {103}},\ \bibinfo {pages} {251} (\bibinfo {year} {1984})}\BibitemShut
  {NoStop}%
\bibitem [{\citenamefont {Andrson}(1966)}]{superfluid_anderson}%
  \BibitemOpen
  \bibfield  {author} {\bibinfo {author} {\bibfnamefont {P.~W.}\ \bibnamefont
  {Andrson}},\ }\bibfield  {title} {\bibinfo {title} {Considerations on the
  flow of superfluid helium},\ }\href
  {https://doi.org/10.1103/RevModPhys.38.298} {\bibfield  {journal} {\bibinfo
  {journal} {Rev. Mod. Phys.}\ }\textbf {\bibinfo {volume} {38}},\ \bibinfo
  {pages} {298} (\bibinfo {year} {1966})}\BibitemShut {NoStop}%
\bibitem [{\citenamefont {Schmitt}(2015)}]{superfluid_book}%
  \BibitemOpen
  \bibfield  {author} {\bibinfo {author} {\bibfnamefont {A.}~\bibnamefont
  {Schmitt}},\ }\href
  {https://doi.org/https://doi.org/10.1007/978-3-319-07947-9} {\emph {\bibinfo
  {title} {Introduction to Superfluidity}}}\ (\bibinfo  {publisher} {Springer
  International Publishing},\ \bibinfo {year} {2015})\BibitemShut {NoStop}%
\bibitem [{\citenamefont {Rempe}\ \emph {et~al.}(1991)\citenamefont {Rempe},
  \citenamefont {Thompson}, \citenamefont {Brecha}, \citenamefont {Lee},\ and\
  \citenamefont {Kimble}}]{cvtqedexp1}%
  \BibitemOpen
  \bibfield  {author} {\bibinfo {author} {\bibfnamefont {G.}~\bibnamefont
  {Rempe}}, \bibinfo {author} {\bibfnamefont {R.~J.}\ \bibnamefont {Thompson}},
  \bibinfo {author} {\bibfnamefont {R.~J.}\ \bibnamefont {Brecha}}, \bibinfo
  {author} {\bibfnamefont {W.~D.}\ \bibnamefont {Lee}},\ and\ \bibinfo {author}
  {\bibfnamefont {H.~J.}\ \bibnamefont {Kimble}},\ }\bibfield  {title}
  {\bibinfo {title} {Optical bistability and photon statistics in cavity
  quantum electrodynamics},\ }\href
  {https://doi.org/10.1103/PhysRevLett.67.1727} {\bibfield  {journal} {\bibinfo
   {journal} {Phys. Rev. Lett.}\ }\textbf {\bibinfo {volume} {67}},\ \bibinfo
  {pages} {1727} (\bibinfo {year} {1991})}\BibitemShut {NoStop}%
\bibitem [{\citenamefont {Thompson}\ \emph {et~al.}(1992)\citenamefont
  {Thompson}, \citenamefont {Rempe},\ and\ \citenamefont
  {Kimble}}]{cvtqedexp2}%
  \BibitemOpen
  \bibfield  {author} {\bibinfo {author} {\bibfnamefont {R.~J.}\ \bibnamefont
  {Thompson}}, \bibinfo {author} {\bibfnamefont {G.}~\bibnamefont {Rempe}},\
  and\ \bibinfo {author} {\bibfnamefont {H.~J.}\ \bibnamefont {Kimble}},\
  }\bibfield  {title} {\bibinfo {title} {Observation of normal-mode splitting
  for an atom in an optical cavity},\ }\href
  {https://doi.org/10.1103/PhysRevLett.68.1132} {\bibfield  {journal} {\bibinfo
   {journal} {Phys. Rev. Lett.}\ }\textbf {\bibinfo {volume} {68}},\ \bibinfo
  {pages} {1132} (\bibinfo {year} {1992})}\BibitemShut {NoStop}%
\bibitem [{\citenamefont {Miller}\ \emph {et~al.}(2005)\citenamefont {Miller},
  \citenamefont {Northup}, \citenamefont {Birnbaum}, \citenamefont {Boca},
  \citenamefont {Boozer},\ and\ \citenamefont {Kimble}}]{cvtqedrev1}%
  \BibitemOpen
  \bibfield  {author} {\bibinfo {author} {\bibfnamefont {R.}~\bibnamefont
  {Miller}}, \bibinfo {author} {\bibfnamefont {T.~E.}\ \bibnamefont {Northup}},
  \bibinfo {author} {\bibfnamefont {K.~M.}\ \bibnamefont {Birnbaum}}, \bibinfo
  {author} {\bibfnamefont {A.}~\bibnamefont {Boca}}, \bibinfo {author}
  {\bibfnamefont {A.~D.}\ \bibnamefont {Boozer}},\ and\ \bibinfo {author}
  {\bibfnamefont {H.~J.}\ \bibnamefont {Kimble}},\ }\bibfield  {title}
  {\bibinfo {title} {Trapped atoms in cavity qed: coupling quantized light and
  matter},\ }\href {https://doi.org/10.1088/0953-4075/38/9/007} {\bibfield
  {journal} {\bibinfo  {journal} {Journal of Physics B: Atomic, Molecular and
  Optical Physics}\ }\textbf {\bibinfo {volume} {38}},\ \bibinfo {pages} {S551}
  (\bibinfo {year} {2005})}\BibitemShut {NoStop}%
\bibitem [{\citenamefont {Walther}\ \emph {et~al.}(2006)\citenamefont
  {Walther}, \citenamefont {Varcoe}, \citenamefont {Englert},\ and\
  \citenamefont {Becker}}]{cvtqedrev2}%
  \BibitemOpen
  \bibfield  {author} {\bibinfo {author} {\bibfnamefont {H.}~\bibnamefont
  {Walther}}, \bibinfo {author} {\bibfnamefont {B.~T.~H.}\ \bibnamefont
  {Varcoe}}, \bibinfo {author} {\bibfnamefont {B.-G.}\ \bibnamefont
  {Englert}},\ and\ \bibinfo {author} {\bibfnamefont {T.}~\bibnamefont
  {Becker}},\ }\bibfield  {title} {\bibinfo {title} {Cavity quantum
  electrodynamics},\ }\href {https://doi.org/10.1088/0034-4885/69/5/r02}
  {\bibfield  {journal} {\bibinfo  {journal} {Reports on Progress in Physics}\
  }\textbf {\bibinfo {volume} {69}},\ \bibinfo {pages} {1325} (\bibinfo {year}
  {2006})}\BibitemShut {NoStop}%
\bibitem [{\citenamefont {Bombelli}\ \emph {et~al.}(1986)\citenamefont
  {Bombelli}, \citenamefont {Koul}, \citenamefont {Lee},\ and\ \citenamefont
  {Sorkin}}]{Bombelli}%
  \BibitemOpen
  \bibfield  {author} {\bibinfo {author} {\bibfnamefont {L.}~\bibnamefont
  {Bombelli}}, \bibinfo {author} {\bibfnamefont {R.~K.}\ \bibnamefont {Koul}},
  \bibinfo {author} {\bibfnamefont {J.}~\bibnamefont {Lee}},\ and\ \bibinfo
  {author} {\bibfnamefont {R.~D.}\ \bibnamefont {Sorkin}},\ }\bibfield  {title}
  {\bibinfo {title} {Quantum source of entropy for black holes},\ }\href
  {https://doi.org/10.1103/PhysRevD.34.373} {\bibfield  {journal} {\bibinfo
  {journal} {Phys. Rev. D}\ }\textbf {\bibinfo {volume} {34}},\ \bibinfo
  {pages} {373} (\bibinfo {year} {1986})}\BibitemShut {NoStop}%
\bibitem [{\citenamefont {Srednicki}(1993)}]{Srednicki}%
  \BibitemOpen
  \bibfield  {author} {\bibinfo {author} {\bibfnamefont {M.}~\bibnamefont
  {Srednicki}},\ }\bibfield  {title} {\bibinfo {title} {Entropy and area},\
  }\href {https://doi.org/10.1103/PhysRevLett.71.666} {\bibfield  {journal}
  {\bibinfo  {journal} {Phys. Rev. Lett.}\ }\textbf {\bibinfo {volume} {71}},\
  \bibinfo {pages} {666} (\bibinfo {year} {1993})}\BibitemShut {NoStop}%
\bibitem [{\citenamefont {Huerta}\ and\ \citenamefont {van~der
  Velde}(2023)}]{Huerta_Velde_2023}%
  \BibitemOpen
  \bibfield  {author} {\bibinfo {author} {\bibfnamefont {M.}~\bibnamefont
  {Huerta}}\ and\ \bibinfo {author} {\bibfnamefont {G.}~\bibnamefont {van~der
  Velde}},\ }\bibfield  {title} {\bibinfo {title} {Modular hamiltonian of the
  scalar in the semi infinite line: dimensional reduction for spherically
  symmetric regions},\ }\href {https://doi.org/10.1007/JHEP06(2023)097}
  {\bibfield  {journal} {\bibinfo  {journal} {Journal of High Energy Physics}\
  }\textbf {\bibinfo {volume} {2023}},\ \bibinfo {pages} {97} (\bibinfo {year}
  {2023})}\BibitemShut {NoStop}%
\bibitem [{\citenamefont {Sarkar}\ \emph {et~al.}(2024)\citenamefont {Sarkar},
  \citenamefont {Moitra},\ and\ \citenamefont {Sensarma}}]{Mrinal_Sl}%
  \BibitemOpen
  \bibfield  {author} {\bibinfo {author} {\bibfnamefont {M.~K.}\ \bibnamefont
  {Sarkar}}, \bibinfo {author} {\bibfnamefont {S.}~\bibnamefont {Moitra}},\
  and\ \bibinfo {author} {\bibfnamefont {R.}~\bibnamefont {Sensarma}},\
  }\bibfield  {title} {\bibinfo {title} {Signature of criticality in angular
  momentum resolved entanglement of scalar fields in $d>1$},\ }\href
  {https://doi.org/10.1103/PhysRevB.110.075128} {\bibfield  {journal} {\bibinfo
   {journal} {Phys. Rev. B}\ }\textbf {\bibinfo {volume} {110}},\ \bibinfo
  {pages} {075128} (\bibinfo {year} {2024})}\BibitemShut {NoStop}%
\bibitem [{\citenamefont {Metlitski}\ \emph {et~al.}(2009)\citenamefont
  {Metlitski}, \citenamefont {Fuertes},\ and\ \citenamefont
  {Sachdev}}]{SubirON}%
  \BibitemOpen
  \bibfield  {author} {\bibinfo {author} {\bibfnamefont {M.~A.}\ \bibnamefont
  {Metlitski}}, \bibinfo {author} {\bibfnamefont {C.~A.}\ \bibnamefont
  {Fuertes}},\ and\ \bibinfo {author} {\bibfnamefont {S.}~\bibnamefont
  {Sachdev}},\ }\bibfield  {title} {\bibinfo {title} {Entanglement entropy in
  the $\uppercase{O(N)}$ model},\ }\href
  {https://doi.org/10.1103/PhysRevB.80.115122} {\bibfield  {journal} {\bibinfo
  {journal} {Phys. Rev. B}\ }\textbf {\bibinfo {volume} {80}},\ \bibinfo
  {pages} {115122} (\bibinfo {year} {2009})}\BibitemShut {NoStop}%
\bibitem [{\citenamefont {Whitsitt}\ \emph {et~al.}(2017)\citenamefont
  {Whitsitt}, \citenamefont {Witczak-Krempa},\ and\ \citenamefont
  {Sachdev}}]{SubirWitczakKrempa}%
  \BibitemOpen
  \bibfield  {author} {\bibinfo {author} {\bibfnamefont {S.}~\bibnamefont
  {Whitsitt}}, \bibinfo {author} {\bibfnamefont {W.}~\bibnamefont
  {Witczak-Krempa}},\ and\ \bibinfo {author} {\bibfnamefont {S.}~\bibnamefont
  {Sachdev}},\ }\bibfield  {title} {\bibinfo {title} {Entanglement entropy of
  large-$\uppercase{N}$ wilson-fisher conformal field theory},\ }\href
  {https://doi.org/10.1103/PhysRevB.95.045148} {\bibfield  {journal} {\bibinfo
  {journal} {Phys. Rev. B}\ }\textbf {\bibinfo {volume} {95}},\ \bibinfo
  {pages} {045148} (\bibinfo {year} {2017})}\BibitemShut {NoStop}%
\bibitem [{\citenamefont {Hertzberg}(2012)}]{Hertzberg_2013}%
  \BibitemOpen
  \bibfield  {author} {\bibinfo {author} {\bibfnamefont {M.~P.}\ \bibnamefont
  {Hertzberg}},\ }\bibfield  {title} {\bibinfo {title} {Entanglement entropy in
  scalar field theory},\ }\href {https://doi.org/10.1088/1751-8113/46/1/015402}
  {\bibfield  {journal} {\bibinfo  {journal} {Journal of Physics A:
  Mathematical and Theoretical}\ }\textbf {\bibinfo {volume} {46}},\ \bibinfo
  {pages} {015402} (\bibinfo {year} {2012})}\BibitemShut {NoStop}%
\bibitem [{\citenamefont {Iso}\ \emph {et~al.}(2021)\citenamefont {Iso},
  \citenamefont {Mori},\ and\ \citenamefont {Sakai}}]{Iso_2021}%
  \BibitemOpen
  \bibfield  {author} {\bibinfo {author} {\bibfnamefont {S.}~\bibnamefont
  {Iso}}, \bibinfo {author} {\bibfnamefont {T.}~\bibnamefont {Mori}},\ and\
  \bibinfo {author} {\bibfnamefont {K.}~\bibnamefont {Sakai}},\ }\bibfield
  {title} {\bibinfo {title} {Entanglement entropy in scalar field theory and
  ${\mathbb{z}}_{M}$ gauge theory on feynman diagrams},\ }\href
  {https://doi.org/10.1103/PhysRevD.103.105010} {\bibfield  {journal} {\bibinfo
   {journal} {Phys. Rev. D}\ }\textbf {\bibinfo {volume} {103}},\ \bibinfo
  {pages} {105010} (\bibinfo {year} {2021})}\BibitemShut {NoStop}%
\bibitem [{\citenamefont {Cotler}\ and\ \citenamefont
  {Mueller}(2016)}]{COTLER_Variational}%
  \BibitemOpen
  \bibfield  {author} {\bibinfo {author} {\bibfnamefont {J.~S.}\ \bibnamefont
  {Cotler}}\ and\ \bibinfo {author} {\bibfnamefont {M.~T.}\ \bibnamefont
  {Mueller}},\ }\bibfield  {title} {\bibinfo {title} {Entanglement entropy and
  variational methods: Interacting scalar fields},\ }\href
  {https://doi.org/https://doi.org/10.1016/j.aop.2015.12.005} {\bibfield
  {journal} {\bibinfo  {journal} {Annals of Physics}\ }\textbf {\bibinfo
  {volume} {365}},\ \bibinfo {pages} {91} (\bibinfo {year} {2016})}\BibitemShut
  {NoStop}%
\bibitem [{\citenamefont {Polkovnikov}\ \emph {et~al.}(2011)\citenamefont
  {Polkovnikov}, \citenamefont {Sengupta}, \citenamefont {Silva},\ and\
  \citenamefont {Vengalattore}}]{RMP_polkovnikov}%
  \BibitemOpen
  \bibfield  {author} {\bibinfo {author} {\bibfnamefont {A.}~\bibnamefont
  {Polkovnikov}}, \bibinfo {author} {\bibfnamefont {K.}~\bibnamefont
  {Sengupta}}, \bibinfo {author} {\bibfnamefont {A.}~\bibnamefont {Silva}},\
  and\ \bibinfo {author} {\bibfnamefont {M.}~\bibnamefont {Vengalattore}},\
  }\bibfield  {title} {\bibinfo {title} {Colloquium: Nonequilibrium dynamics of
  closed interacting quantum systems},\ }\href
  {https://doi.org/10.1103/RevModPhys.83.863} {\bibfield  {journal} {\bibinfo
  {journal} {Rev. Mod. Phys.}\ }\textbf {\bibinfo {volume} {83}},\ \bibinfo
  {pages} {863} (\bibinfo {year} {2011})}\BibitemShut {NoStop}%
\bibitem [{\citenamefont {Moessner}\ and\ \citenamefont
  {Sondhi}(2017)}]{Moessner2017}%
  \BibitemOpen
  \bibfield  {author} {\bibinfo {author} {\bibfnamefont {R.}~\bibnamefont
  {Moessner}}\ and\ \bibinfo {author} {\bibfnamefont {S.~L.}\ \bibnamefont
  {Sondhi}},\ }\bibfield  {title} {\bibinfo {title} {Equilibration and order in
  quantum floquet matter},\ }\href {https://doi.org/10.1038/nphys4106}
  {\bibfield  {journal} {\bibinfo  {journal} {Nature Physics}\ }\textbf
  {\bibinfo {volume} {13}},\ \bibinfo {pages} {424} (\bibinfo {year}
  {2017})}\BibitemShut {NoStop}%
\bibitem [{\citenamefont {Pretko}(2018)}]{Pretko_dipole}%
  \BibitemOpen
  \bibfield  {author} {\bibinfo {author} {\bibfnamefont {M.}~\bibnamefont
  {Pretko}},\ }\bibfield  {title} {\bibinfo {title} {The fracton gauge
  principle},\ }\href {https://doi.org/10.1103/PhysRevB.98.115134} {\bibfield
  {journal} {\bibinfo  {journal} {Phys. Rev. B}\ }\textbf {\bibinfo {volume}
  {98}},\ \bibinfo {pages} {115134} (\bibinfo {year} {2018})}\BibitemShut
  {NoStop}%
\bibitem [{\citenamefont {Islam}\ \emph {et~al.}(2023)\citenamefont {Islam},
  \citenamefont {Sengupta},\ and\ \citenamefont {Sensarma}}]{Mursalin_dipole}%
  \BibitemOpen
  \bibfield  {author} {\bibinfo {author} {\bibfnamefont {M.~M.}\ \bibnamefont
  {Islam}}, \bibinfo {author} {\bibfnamefont {K.}~\bibnamefont {Sengupta}},\
  and\ \bibinfo {author} {\bibfnamefont {R.}~\bibnamefont {Sensarma}},\
  }\bibfield  {title} {\bibinfo {title} {Nonequilibrium dynamics of bosons with
  dipole symmetry: Large-$n$ keldysh approach},\ }\href
  {https://doi.org/10.1103/PhysRevB.108.214314} {\bibfield  {journal} {\bibinfo
   {journal} {Phys. Rev. B}\ }\textbf {\bibinfo {volume} {108}},\ \bibinfo
  {pages} {214314} (\bibinfo {year} {2023})}\BibitemShut {NoStop}%
\bibitem [{\citenamefont {Cahill}\ and\ \citenamefont
  {Glauber}(1969)}]{Cahill_Glauber_1969}%
  \BibitemOpen
  \bibfield  {author} {\bibinfo {author} {\bibfnamefont {K.~E.}\ \bibnamefont
  {Cahill}}\ and\ \bibinfo {author} {\bibfnamefont {R.~J.}\ \bibnamefont
  {Glauber}},\ }\bibfield  {title} {\bibinfo {title} {Density operators and
  quasiprobability distributions},\ }\href
  {https://doi.org/10.1103/PhysRev.177.1882} {\bibfield  {journal} {\bibinfo
  {journal} {Phys. Rev.}\ }\textbf {\bibinfo {volume} {177}},\ \bibinfo {pages}
  {1882} (\bibinfo {year} {1969})}\BibitemShut {NoStop}%
\bibitem [{\citenamefont {Chakraborty}\ and\ \citenamefont
  {Sensarma}(2021{\natexlab{a}})}]{AhanaPRL}%
  \BibitemOpen
  \bibfield  {author} {\bibinfo {author} {\bibfnamefont {A.}~\bibnamefont
  {Chakraborty}}\ and\ \bibinfo {author} {\bibfnamefont {R.}~\bibnamefont
  {Sensarma}},\ }\bibfield  {title} {\bibinfo {title} {Nonequilibrium dynamics
  of renyi entropy for bosonic many-particle systems},\ }\href
  {https://doi.org/10.1103/PhysRevLett.127.200603} {\bibfield  {journal}
  {\bibinfo  {journal} {Phys. Rev. Lett.}\ }\textbf {\bibinfo {volume} {127}},\
  \bibinfo {pages} {200603} (\bibinfo {year} {2021}{\natexlab{a}})}\BibitemShut
  {NoStop}%
\bibitem [{\citenamefont {Chakraborty}\ and\ \citenamefont
  {Sensarma}(2021{\natexlab{b}})}]{AhanaPRA}%
  \BibitemOpen
  \bibfield  {author} {\bibinfo {author} {\bibfnamefont {A.}~\bibnamefont
  {Chakraborty}}\ and\ \bibinfo {author} {\bibfnamefont {R.}~\bibnamefont
  {Sensarma}},\ }\bibfield  {title} {\bibinfo {title} {Renyi entropy of
  interacting thermal bosons in the large-$n$ approximation},\ }\href
  {https://doi.org/10.1103/PhysRevA.104.032408} {\bibfield  {journal} {\bibinfo
   {journal} {Phys. Rev. A}\ }\textbf {\bibinfo {volume} {104}},\ \bibinfo
  {pages} {032408} (\bibinfo {year} {2021}{\natexlab{b}})}\BibitemShut
  {NoStop}%
\bibitem [{\citenamefont {Moitra}\ and\ \citenamefont
  {Sensarma}(2020)}]{moitra2020entanglement}%
  \BibitemOpen
  \bibfield  {author} {\bibinfo {author} {\bibfnamefont {S.}~\bibnamefont
  {Moitra}}\ and\ \bibinfo {author} {\bibfnamefont {R.}~\bibnamefont
  {Sensarma}},\ }\bibfield  {title} {\bibinfo {title} {Entanglement entropy of
  fermions from wigner functions: Excited states and open quantum systems},\
  }\href {https://doi.org/10.1103/PhysRevB.102.184306} {\bibfield  {journal}
  {\bibinfo  {journal} {Phys. Rev. B}\ }\textbf {\bibinfo {volume} {102}},\
  \bibinfo {pages} {184306} (\bibinfo {year} {2020})}\BibitemShut {NoStop}%
\bibitem [{\citenamefont {Moitra}\ and\ \citenamefont
  {Sensarma}(2023)}]{Saranyo_Building_Entanglement}%
  \BibitemOpen
  \bibfield  {author} {\bibinfo {author} {\bibfnamefont {S.}~\bibnamefont
  {Moitra}}\ and\ \bibinfo {author} {\bibfnamefont {R.}~\bibnamefont
  {Sensarma}},\ }\bibfield  {title} {\bibinfo {title} {Building entanglement
  entropy out of correlation functions for interacting fermions},\ }\href
  {https://doi.org/10.1103/PhysRevB.108.174309} {\bibfield  {journal} {\bibinfo
   {journal} {Phys. Rev. B}\ }\textbf {\bibinfo {volume} {108}},\ \bibinfo
  {pages} {174309} (\bibinfo {year} {2023})}\BibitemShut {NoStop}%
\bibitem [{\citenamefont {Cotler}\ \emph {et~al.}(2016)\citenamefont {Cotler},
  \citenamefont {Hertzberg}, \citenamefont {Mezei},\ and\ \citenamefont
  {Mueller}}]{Cotler2016}%
  \BibitemOpen
  \bibfield  {author} {\bibinfo {author} {\bibfnamefont {J.~S.}\ \bibnamefont
  {Cotler}}, \bibinfo {author} {\bibfnamefont {M.~P.}\ \bibnamefont
  {Hertzberg}}, \bibinfo {author} {\bibfnamefont {M.}~\bibnamefont {Mezei}},\
  and\ \bibinfo {author} {\bibfnamefont {M.~T.}\ \bibnamefont {Mueller}},\
  }\bibfield  {title} {\bibinfo {title} {Entanglement growth after a global
  quench in free scalar field theory},\ }\href
  {https://doi.org/10.1007/JHEP11(2016)166} {\bibfield  {journal} {\bibinfo
  {journal} {Journal of High Energy Physics}\ }\textbf {\bibinfo {volume}
  {2016}},\ \bibinfo {pages} {166} (\bibinfo {year} {2016})}\BibitemShut
  {NoStop}%
\bibitem [{\citenamefont {Holzhey}\ \emph {et~al.}(1994)\citenamefont
  {Holzhey}, \citenamefont {Larsen},\ and\ \citenamefont
  {Wilczek}}]{HOLZHEY1994443}%
  \BibitemOpen
  \bibfield  {author} {\bibinfo {author} {\bibfnamefont {C.}~\bibnamefont
  {Holzhey}}, \bibinfo {author} {\bibfnamefont {F.}~\bibnamefont {Larsen}},\
  and\ \bibinfo {author} {\bibfnamefont {F.}~\bibnamefont {Wilczek}},\
  }\bibfield  {title} {\bibinfo {title} {Geometric and renormalized entropy in
  conformal field theory},\ }\href
  {https://doi.org/https://doi.org/10.1016/0550-3213(94)90402-2} {\bibfield
  {journal} {\bibinfo  {journal} {Nuclear Physics B}\ }\textbf {\bibinfo
  {volume} {424}},\ \bibinfo {pages} {443} (\bibinfo {year}
  {1994})}\BibitemShut {NoStop}%
\bibitem [{\citenamefont {Anisimov}\ \emph {et~al.}(2009)\citenamefont
  {Anisimov}, \citenamefont {Buchmüller}, \citenamefont {Drewes},\ and\
  \citenamefont {Mendizabal}}]{noneqscalar_anisimov}%
  \BibitemOpen
  \bibfield  {author} {\bibinfo {author} {\bibfnamefont {A.}~\bibnamefont
  {Anisimov}}, \bibinfo {author} {\bibfnamefont {W.}~\bibnamefont
  {Buchmüller}}, \bibinfo {author} {\bibfnamefont {M.}~\bibnamefont
  {Drewes}},\ and\ \bibinfo {author} {\bibfnamefont {S.}~\bibnamefont
  {Mendizabal}},\ }\bibfield  {title} {\bibinfo {title} {Nonequilibrium
  dynamics of scalar fields in a thermal bath},\ }\href
  {https://doi.org/https://doi.org/10.1016/j.aop.2009.01.001} {\bibfield
  {journal} {\bibinfo  {journal} {Annals of Physics}\ }\textbf {\bibinfo
  {volume} {324}},\ \bibinfo {pages} {1234} (\bibinfo {year}
  {2009})}\BibitemShut {NoStop}%
\bibitem [{\citenamefont {Islam}\ and\ \citenamefont
  {Sensarma}(2022)}]{Mursalin2022}%
  \BibitemOpen
  \bibfield  {author} {\bibinfo {author} {\bibfnamefont {M.~M.}\ \bibnamefont
  {Islam}}\ and\ \bibinfo {author} {\bibfnamefont {R.}~\bibnamefont
  {Sensarma}},\ }\bibfield  {title} {\bibinfo {title} {Nonequilibrium scalar
  field dynamics starting from fock states: Absence of thermalization in
  one-dimensional phonons coupled to fermions},\ }\href
  {https://doi.org/10.1103/PhysRevB.106.024306} {\bibfield  {journal} {\bibinfo
   {journal} {Phys. Rev. B}\ }\textbf {\bibinfo {volume} {106}},\ \bibinfo
  {pages} {024306} (\bibinfo {year} {2022})}\BibitemShut {NoStop}%
\bibitem [{\citenamefont {Laguna}\ and\ \citenamefont
  {Zurek}(1997)}]{LagunaPabloZurek}%
  \BibitemOpen
  \bibfield  {author} {\bibinfo {author} {\bibfnamefont {P.}~\bibnamefont
  {Laguna}}\ and\ \bibinfo {author} {\bibfnamefont {W.~H.}\ \bibnamefont
  {Zurek}},\ }\bibfield  {title} {\bibinfo {title} {Density of kinks after a
  quench: When symmetry breaks, how big are the pieces?},\ }\href
  {https://doi.org/10.1103/PhysRevLett.78.2519} {\bibfield  {journal} {\bibinfo
   {journal} {Phys. Rev. Lett.}\ }\textbf {\bibinfo {volume} {78}},\ \bibinfo
  {pages} {2519} (\bibinfo {year} {1997})}\BibitemShut {NoStop}%
\bibitem [{\citenamefont {Halperin}\ and\ \citenamefont
  {Hohenberg}(1969)}]{Halparin}%
  \BibitemOpen
  \bibfield  {author} {\bibinfo {author} {\bibfnamefont {B.~I.}\ \bibnamefont
  {Halperin}}\ and\ \bibinfo {author} {\bibfnamefont {P.~C.}\ \bibnamefont
  {Hohenberg}},\ }\bibfield  {title} {\bibinfo {title} {Scaling laws for
  dynamic critical phenomena},\ }\href
  {https://doi.org/10.1103/PhysRev.177.952} {\bibfield  {journal} {\bibinfo
  {journal} {Phys. Rev.}\ }\textbf {\bibinfo {volume} {177}},\ \bibinfo {pages}
  {952} (\bibinfo {year} {1969})}\BibitemShut {NoStop}%
\bibitem [{\citenamefont {Weiher}\ \emph {et~al.}(2019)\citenamefont {Weiher},
  \citenamefont {Agudelo},\ and\ \citenamefont {Bohmann}}]{Weiher}%
  \BibitemOpen
  \bibfield  {author} {\bibinfo {author} {\bibfnamefont {K.}~\bibnamefont
  {Weiher}}, \bibinfo {author} {\bibfnamefont {E.}~\bibnamefont {Agudelo}},\
  and\ \bibinfo {author} {\bibfnamefont {M.}~\bibnamefont {Bohmann}},\
  }\bibfield  {title} {\bibinfo {title} {Conditional nonclassical field
  generation in cavity qed},\ }\href
  {https://doi.org/10.1103/PhysRevA.100.043812} {\bibfield  {journal} {\bibinfo
   {journal} {Phys. Rev. A}\ }\textbf {\bibinfo {volume} {100}},\ \bibinfo
  {pages} {043812} (\bibinfo {year} {2019})}\BibitemShut {NoStop}%
\bibitem [{\citenamefont {Bao}\ \emph {et~al.}(2019)\citenamefont {Bao},
  \citenamefont {Zhu}, \citenamefont {Yang},\ and\ \citenamefont
  {Agarwal}}]{cvtqwed_sqzlit3}%
  \BibitemOpen
  \bibfield  {author} {\bibinfo {author} {\bibfnamefont {D.~Q.}\ \bibnamefont
  {Bao}}, \bibinfo {author} {\bibfnamefont {C.~J.}\ \bibnamefont {Zhu}},
  \bibinfo {author} {\bibfnamefont {Y.~P.}\ \bibnamefont {Yang}},\ and\
  \bibinfo {author} {\bibfnamefont {G.~S.}\ \bibnamefont {Agarwal}},\
  }\bibfield  {title} {\bibinfo {title} {Sensing single atoms in a cavity using
  a broadband squeezed light},\ }\href {https://doi.org/10.1364/OE.27.015540}
  {\bibfield  {journal} {\bibinfo  {journal} {Opt. Express}\ }\textbf {\bibinfo
  {volume} {27}},\ \bibinfo {pages} {15540} (\bibinfo {year}
  {2019})}\BibitemShut {NoStop}%
\bibitem [{\citenamefont {Alba}\ and\ \citenamefont
  {Calabrese}(2017)}]{Alba_calabrese}%
  \BibitemOpen
  \bibfield  {author} {\bibinfo {author} {\bibfnamefont {V.}~\bibnamefont
  {Alba}}\ and\ \bibinfo {author} {\bibfnamefont {P.}~\bibnamefont
  {Calabrese}},\ }\bibfield  {title} {\bibinfo {title} {Entanglement and
  thermodynamics after a quantum quench in integrable systems},\ }\href
  {https://doi.org/10.1073/pnas.1703516114} {\bibfield  {journal} {\bibinfo
  {journal} {Proceedings of the National Academy of Sciences}\ }\textbf
  {\bibinfo {volume} {114}},\ \bibinfo {pages} {7947} (\bibinfo {year}
  {2017})},\ \Eprint
  {https://arxiv.org/abs/https://www.pnas.org/doi/pdf/10.1073/pnas.1703516114}
  {https://www.pnas.org/doi/pdf/10.1073/pnas.1703516114} \BibitemShut {NoStop}%
\bibitem [{\citenamefont {Chakraborty}\ and\ \citenamefont
  {Sensarma}(2018)}]{Power_law_Ahana}%
  \BibitemOpen
  \bibfield  {author} {\bibinfo {author} {\bibfnamefont {A.}~\bibnamefont
  {Chakraborty}}\ and\ \bibinfo {author} {\bibfnamefont {R.}~\bibnamefont
  {Sensarma}},\ }\bibfield  {title} {\bibinfo {title} {Power-law tails and
  non-markovian dynamics in open quantum systems: An exact solution from
  keldysh field theory},\ }\href {https://doi.org/10.1103/PhysRevB.97.104306}
  {\bibfield  {journal} {\bibinfo  {journal} {Phys. Rev. B}\ }\textbf {\bibinfo
  {volume} {97}},\ \bibinfo {pages} {104306} (\bibinfo {year}
  {2018})}\BibitemShut {NoStop}%
\bibitem [{\citenamefont {Caputa}\ \emph {et~al.}(2017)\citenamefont {Caputa},
  \citenamefont {Das}, \citenamefont {Nozaki},\ and\ \citenamefont
  {Tomiya}}]{Sumit_Das_quench}%
  \BibitemOpen
  \bibfield  {author} {\bibinfo {author} {\bibfnamefont {P.}~\bibnamefont
  {Caputa}}, \bibinfo {author} {\bibfnamefont {S.~R.}\ \bibnamefont {Das}},
  \bibinfo {author} {\bibfnamefont {M.}~\bibnamefont {Nozaki}},\ and\ \bibinfo
  {author} {\bibfnamefont {A.}~\bibnamefont {Tomiya}},\ }\bibfield  {title}
  {\bibinfo {title} {Quantum quench and scaling of entanglement entropy},\
  }\href {https://doi.org/https://doi.org/10.1016/j.physletb.2017.06.017}
  {\bibfield  {journal} {\bibinfo  {journal} {Physics Letters B}\ }\textbf
  {\bibinfo {volume} {772}},\ \bibinfo {pages} {53} (\bibinfo {year}
  {2017})}\BibitemShut {NoStop}%
\bibitem [{\citenamefont {Banerjee}\ \emph {et~al.}(2020)\citenamefont
  {Banerjee}, \citenamefont {Gaikwad}, \citenamefont {Kaushal},\ and\
  \citenamefont {Mandal}}]{banerjee2020quantum}%
  \BibitemOpen
  \bibfield  {author} {\bibinfo {author} {\bibfnamefont {P.}~\bibnamefont
  {Banerjee}}, \bibinfo {author} {\bibfnamefont {A.}~\bibnamefont {Gaikwad}},
  \bibinfo {author} {\bibfnamefont {A.}~\bibnamefont {Kaushal}},\ and\ \bibinfo
  {author} {\bibfnamefont {G.}~\bibnamefont {Mandal}},\ }\bibfield  {title}
  {\bibinfo {title} {Quantum quench and thermalization to gge in arbitrary
  dimensions and the odd-even effect},\ }\href@noop {} {\bibfield  {journal}
  {\bibinfo  {journal} {Journal of High Energy Physics}\ }\textbf {\bibinfo
  {volume} {2020}},\ \bibinfo {pages} {1} (\bibinfo {year} {2020})}\BibitemShut
  {NoStop}%
\end{thebibliography}%
		
	\end{document}